# On the parallels between Paxos and Raft, and how to port optimizations (Extended Version)

ZHAOGUO WANG[†‡], CHANGGENG ZHAO[∗], SHUAI MU[⋄], HAIBO CHEN[†‡], JINYANG LI[∗], † Shanghai Key Laboratory of Scalable Computing and Systems, Shanghai Jiao Tong University, ‡ Institute of Parallel and Distributed Systems, Shanghai Jiao Tong University, ∗ Department of Computer Science, New York University, and ⋄ Department of Computer Science, Stony Brook University

In recent years, Raft has overtaken Paxos as the consensus algorithm of choice. [53] While many have pointed out similarities between the two protocols, no one has formally mapped out their relationships. In this paper, we show how Raft and Paxos are formally related despite their surface differences. Based on the formal mapping between the two protocols, we show how to automatically port a certain class of optimizations from Paxos to Raft with guaranteed correctness. As case studies, we port and evaluate two optimizations, Mencius and Paxos Quorum Lease to Raft.

Contents

# 1 INTRODUCTION

Consensus protocols enable servers to reach agreement on the sequence of operations to execute despite the failure of some servers and arbitrary network delays. Classic Paxos is one of the oldest and most well-studied consensus protocols. However, in recent years, Raft has gradually overtaken Paxos as the consensus protocol of choice, esp. in the industry. Many researchers have observed that Raft and Paxos bear certain similarities. However, no one has shown how the two protocols are related in the formal sense. In fact, does such a formal relationship exist?

While it may seem like a pedantic endeavor, investigating a formal mapping between Raft and Paxos is meaningful for two reasons. First, making the connection between Raft and Paxos helps deepen our understanding of both protocols. In particular, it allows us to articulate what design decisions have made Raft more understandable or more efficient than Paxos. Second, Paxos is not an isolated protocol but consists of a large family of variants and optimizations as a result of almost two decades of research [38, 47, 19, 49, 45, 54, 53, 5]. These Paxos variants range from reducing latency for wide-area operation, balancing replica load, optimizing for mostly-conflict-free workload, to Byzantine fault tolerance. Knowing how Raft relates to Paxos allows one to port some of these optimizations to Raft without having to reinvent the wheel.

In this paper, we attempt to make a formal connection between Raft and Paxos using refinement mapping. We show that, beyond the broad stroke similarities between the two protocols, Raft differs from Paxos in several interesting details, such as allowing the follower to erase extra entries if its log is longer than the leader, avoiding modifying the persisted proposals. Unfortunately, these surface differences between the two protocols prevent direct refinement mapping between them. Therefore, we present a variant of Raft, called Raft*, which is a refinement of Paxos by removing these surface differences.

We use the refinement mapping between Raft* and Paxos to port existing ideas in the Paxos literature to the world of Raft. Specifically, we develop an automatic porting method which is able to port a certain class of Paxos optimizations to Raft*. The specific class of optimizations that allow automatic porting are those that do not mutate the original state in Paxos. For these optimizations, we derive the set of rules for applying them to Raft*, such that the resulting protocol is automatically guaranteed to refine Raft* and the Paxos optimization, thus remains correct. As Raft* is very close to Raft, the derived algorithm contains all Raft properties and is improved by the Paxos optimization. As case studies, we choose two published Paxos optimization, Mencius and Paxos Quorum Lease, each of which improves one or more aspects of Paxos in terms of load-balancing and latency. We have ported these two protocols to Raft*.

We evaluate the performance benefits of our Raft* optimizations on Amazon AWS in a setup where data is replicated across several geographically separated data centers. For each optimization, we show that the Raft variant has similar benefits to its Paxos counterpart in the literature.

To summarize, we make the following contributions:

- We reveal the relation between Raft and Paxos by connecting them with Raft close variant (Section 3).
- We define the problem of automated porting of optimizations across protocols and develop a methodology for automatically porting a restricted class of optimizations (Section 4)
- Porting Mencius and Quorum Lease from Paxos to Raft and providing experimental evaluation of the optimized versions of Raft protocols (Section 5)

**Limitations.** As the first-timer on the problem of optimization porting, this paper has the following limitations. First, the method proposed can only port a restricted class of optimizations. Second, the correctness of the derived algorithm is only guaranteed in term of specification. The programmer needs manual effort to ensure the correctness of the implementation. Last, this paper found the equivalence between Paxos and a Raft variant, Raft*. Thus, we automatically port Paxos optimization to Raft* instead of Raft. However, Raft* is lightly different with Raft, we believe port optimization from Raft* to Raft is trivial.

## 2  OVERVIEW

### 2.1  background

```
1   function Phase1a(s):
2     s.ballot = s.ballot + 1
3     unchosen = smallest unchosen instance
          id
4     s sends <''prepare'', s.ballot, unchosen> to
          all
5
6   function Phase1b(s):
7     if s receives <''prepare'', b, unchosen>
8        && b > s.ballot
9     then
10      s.ballot = b
11      s.phase1Succeeded = false
12      reply <''prepareOK'', s.ballot,
13                    instances with id ≥
                          unchosen>
14
15  function Phase1Succeed(s):
16    if s receives <''prepareOK'', b, instances>
17       from f + 1 acceptors with the same b
18       && b == s.ballot
19    then
20      start = smallest unchosen instance id
21      end = largest id of received
             instances
22      for i in start ... end
23        s.instances[i]= safeEntry(received
24                         instances with id
                                 i)
25      s.phase1Succeeded = true
```

(a) Phase 1

```
function Phase2a(s, i, v):
  if s.phase1Succeeded
     && (s.instances[i].val == v
         || s.instances[i] == Empty)
  then
    send <''accept'', i, v, s.ballot> to all

function Phase2b(s):
  if s receives <''accept'', i, v, b> && b ≥
      s.ballot
  then
    if b > s.ballot
    then
      s.phase1Succeeded = false
    s.ballot = b
    s.instances[i].bal = b
    s.instances[i].val = v
    reply <''acceptOK'', i, v, b>

function Learn(s):
  if s receives same <''acceptOK'', i, v, b>
     from f + 1 acceptors
  then
    s.instances[i].val = v
    s.instances[i].bal = b
    add s.instances[i] to s.chosenSet
```

(b) Phase 2

Fig. 1. MultiPaxos

**Paxos [40].** Paxos solves the consensus problem in log replication using a bottom-up approach by first solving the single-decree consensus problem that allows servers to agree on a single value. In single-decree Paxos, each server plays two roles, either as a proposer or an acceptor. Proposers make proposals, and acceptors vote for proposals. Servers try to reach consensus via a two-phase protocol.

In the first phase, a server picks a globally unique proposal number and sends a Prepare RPC to every server. The proposal number is often called the ballot in the Paxos literature. A server receiving a Prepare replies success if it has not seen a higher ballot. In its reply, the server will also include the highest ballot it has ever accepted, or *null* if it has never accepted any ballot. If the proposing server receives at least a majority ($f + 1$) of successful replies, it goes into the second phase, otherwise, it retries with a higher ballot.

In the second phase, the server picks the value for its ballot and sends it in an Accept RPC to every server. This can be any value (usually the operation the server wants to initiate) if the replies do not contain any previously accepted value; Otherwise it must be the value (usually an operation some other server trying to initiate) with the highest ballot in the replies. A server receiving this Accept RPC will accept this value and reply success if it has not seen a higher ballot. If the proposing server receives a majority of successful replies, it considers this value is chosen (or equivalently, *committed*). It could notify other servers of the committed operation immediately or piggyback this information with subsequent communication.

MultiPaxos builds upon single-decree Paxos to agree on a sequence of operations. In particular, MultiPaxos tries to agree on the operation for each position in the sequence using a separate instance of Paxos. MultiPaxos also optimizes performance by batching the phase-1 of many single-decree instances and allowing concurrent instances. More will be discussed in Section 3. Figure 1 gives the pseudocode of MultiPaxos. In the following paragraphs, we use the term Paxos to refer to the multi-decree version of Paxos and use the term single-decree Paxos to refer to the single-decree version.

**Raft [57].** Unlike Paxos' bottom-up approach, Raft solves the consensus problem in log replication in a top-down manner, without decomposing it to single-decree consensus.

The Raft protocol consists of two parts: electing a leader and replicating of log entries by the elected leader. Each server maintains a term number that monotonically increases. For leader election, a candidate server increments its term number and sends RequestVote RPCs to all servers to collect votes for itself to become a leader. Elections are ordered by their corresponding *term* numbers and a node rejects RequestVote if it has already processed a request with a higher term or the same term from a different candidate. Raft also adds another restriction to leader election: a node rejects RequestVote if its log is more recent than the sender's log. A candidate becomes the elected leader for this term if it receives a majority quorum of successful votes on its RequestVote RPCs.

The elected leader batches client operations and replicates them to all other servers (called followers) using the AppendEntries RPC. A server rejects the AppendEntries request if it has seen a higher term than the sender. The AppendEntries RPC also lets the receiver synchronize its log with the sender: the receiver catches up if it is missing entries, and it erases extraneous entries not found in the sender's log. The leader considers the operations in AppendEntries committed if a majority quorum of servers successfully replies. Figure 2 shows the pseudocode of Raft, for now, we ignore the code in blue.

## 2.2 Our approach overview

At high-level, the Raft protocol bears many similarities to MultiPaxos. Both protocols have two phases. Raft's RequestVote corresponds to Phase1a in MultiPaxos. Both RequestVote and Phase1a are considered successful if a majority of ok replies are received. Afterward, Raft uses AppendEntries to replicate operations to other servers, similar to how MultiPaxos uses Accept in the second phase to disseminate an operation associated with a specific Paxos instance. In both protocols, a server rejects the AppendEntries/Accept request if it has seen a higher term/ballot and the operation is considered committed only when a majority of servers return ok to the leader/proposer.

Given these similarities, there are existing works to compare these two protocols [10, 33, 13]. Unfortunately, none of these discussion leads to a formalized mapping of these two protocols. We use refinement mapping [1] to formally capture the connection between Paxos and Raft. Refinement mapping is commonly used to prove that a lower-level specification ($S_L$) correctly implements a higher-level one ($S_H$). In order for $S_L$ to refine $S_H$, we must show that each state in $S_L$'s state space can be mapped to some state in $S_H$'s state space such that any state transition sequence allowed by $S_L$ corresponds to a valid state transition sequence in $S_H$ under the mapping. There exist literature [66] comparing

```
1  function RequestVote(s):
2    s.currentTerm = s.currentTerm + 1
3    s sends <''requestVote'', s.currentTerm,
           s.lastIndex,
4            s.log[lastIndex].term> to all
5
6  function ReceiveVote(s):
7    if s receives <''requestVote'', t, lIndex,
           lTerm>
8         && t > s.currentTerm
9         && s.log[lIndex].term < lTerm
10        || (s.log[lIndex].term = lTerm
11            && s.lastIndex ≥ lIndex)
12   then
13     s.currentTerm = t
14     s.isLeader = false
15     extraEnts = non-empty entries in s.log after
           lIndex
16     reply <''requestVoteOK'', s.currentTerm,
           extraEnts>
17
18 function BecomeLeader(s):
19   if s receives <''requestVoteOK'', t, ents> from
           f + 1
20      acceptors with same t && t == s.
           currentTerm
21   then
22     max = largest index of recieved
           entries
23     for i in lastIndex + 1 ... max
24       e = safeEntry(recieved entries of index i)
25       s.log[i].bal = currentTerm
26       s.log[i].term = currentTerm
27       s.log[i].val = e.val
28     s.isLeader = true
29     s.lastIndex = max
30
31
32
33
34
              (a) Phase 1
```

```
function AppendEntries(s, i, vals, prev):
  if s.isLeader && i == s.lastIndex + 1 then
    for each v in vals
      s.log[s.lastIndex+1].val = v
      s.log[s.lastIndex+1].term = s.
           currentTerm
      s.lastIndex = s.lastIndex + 1
    for i in prev + 1 ... lastIndex
      s.log[i].bal = currentTerm
    ents = s.log entries after prev
    pTerm = s.log[prev].term
    send <''append'', s.currentTerm, prev, pTerm,
           ents, s.commitIndex> to all

function RecieveAppend(s):
  if s receives <''append'', t, prev, pTerm,
           ents, commit>
    && t ≥ s.currentTerm
    && s.log[prev].term == pTerm
    && s.lastIndex ≤ prev + length(ents)
  then
    if t > s.currentTerm then
      s.isLeader = false
      s.currentTerm = t
    s.lastIndex = prev + length(ents)
    s.commitIndex = max(commit, s.
           commitIndex)
    replace entries after s.log[prev]
           with ents
    reply <''appendOK'', s.currentTerm,
           s.lastIndex>

function LeaderLearn(s):
  if s receives <''appendOK'', term, index> from
      f acceptors with the same term
    && s.isLeader
    && s.term == s.currentTerm
  then
    minIndex = minimal received index
    s.commitIndex = max(s.commitIndex,
           minindex)

              (b) Phase 2
```

Fig. 2. Raft*

consensus algorithms (Paxos, VR ,and ZAB) based on refinement mapping to help researchers better understand existing works and close the gaps between technical ancestors. However, this work maps these algorithms to high-level abstractions instead of directly relating them. This paper modifies Raft slightly to create a close invariant, called Raft*, for which a refinement mapping to exists. With Raft*, we are able to show the equivalence between Paxos and Raft.

Furthermore, given the correspondences, we develop a method to port Paxos optimizations to Raft* automatically. The intuition of the porting algorithm is, given a Paxos's optimization, if it only reads the Paxos original variables (e.g., ballot), but never mutate them. Then we can port the optimization to Raft* by replacing Paxos variables with Raft* variables according to the refinement mapping. For example, considering the checkpoint process in Paxos, it needs to checkpoint both system state and last applied instance id. According to the refinement mapping, the instance id is mapped to the log index. Thus, when porting the checkpoint mechanism to Raft*, we can replace the last applied instance id with the last applied index in the log without considering the precise semantics.

## 3 CONNECT RAFT TO PAXOS

**Why Raft cannot be mapped to Paxos directly.** There are two reasons. First, Raft forces all servers that accept the leader's AppendEntries to match the leader's log. Therefore, if a follower's log is longer than that of the leader, the follower will erase the extra entries. When mapped to MultiPaxos, such an "erasing" step would correspond to a server deleting a previously accepted value for some Paxos instance, a state transition that would never happen in MultiPaxos. The erasing step is safe because Raft commits log indexes in order so its phase-1 exchange can guarantee that the leader's log contains all potentially committed entries. By contrast, MultiPaxos commits instances out of order. Thus MultiPaxos' proposing server must fetch safe values for different uncommitted instances from other servers and never erase (but only overwrite) accepted values at other servers. Second, the term number kept with each log entry in Raft cannot be mapped to the accepted ballot number kept with each instance in MultiPaxos. This is because the leader in Raft never modifies its existing log entries. As a result, a newly elected leader at term $t$ would replicate a previously uncommitted log entry with term $t' < t$ without any change. Such a behavior has no equivalent in MultiPaxos. The proposing server in MultiPaxos always over-writes the accepted instance's ballot number with its current ballot number. Not changing the log entry's term number turns out to have subtle correctness implications such that the Raft paper has to add an extra rule to prevent the loss of committed values ([57] Section 5.4.2).

**Raft*.** We modify Raft slightly to create a close variant, called Raft*, for which a refinement mapping to Paxos exists. Figure 2 shows the specification of Raft* in pseudocode. Raft* is identical to Raft, except for two introduced differences, based on the two reasons for why Raft cannot be shown to refine Paxos. First, when responding ok to a candidate's RequestVote, a server includes all the extra entries not present in the candidate's log in its reply (line 14-15). The leader chooses the safe values among its majority quorum of replies to extend its log (line 22-26). An acceptor rejects leader's append request if its log is longer than leaders (Figure 2.b line 16). Second, a ballot field is added to each entry. On appending a new entry, Raft* will change all entries' ballot to be the new entry's term. (Figure 2.b line 6-7 and line 25).

**Refining Paxos with Raft*** Figure 3 gives the mapping between Raft* and Paxos. Most of them are obvious, and we illustrate unobvious ones here. First, for most message pairs, the message in Raft* does not have the same content with its counterpart in Paxos, as it is not necessary. For example, requestVote attaches lastIndex and lastTerm instead of the smallest id of unchosen instance. This is because, with log matching property, using the lastTerm is enough to detect if every entry in a log is more up-to-date. Second, Raft*'s leader directly appends entries into the log in AppendEntries and BecomeLeader. This can be considered as implicitly sending an append to itself, then receiving an appendOk from itself. Both explicit and implicit append/appendOk can imply accept/acceptOk. Last, a Raft*'s function may imply multiple functions in Paxos. For example, AppendEntries implies both Phase2a and Phase2b: the leader first implicitly accept the entry, then send it to others. The formal specification and the formal proof of the refinement mapping can be found in [9].

## 4 A METHOD FOR PORTING OPTIMIZATION

In this section, we show how to automatically port Paxos optimizations to Raft* by leveraging their refinement mapping. First, we define the porting problem formally(??): Given two $\mathcal{A}$ and $\mathcal{B}$, if $\mathcal{B}$ is a refinement of $\mathcal{A}$, how could we automatically adapt $\mathcal{A}$'s optimization (i.e. $\mathcal{A}^\Delta$) to also improve $\mathcal{B}$. To make the problem trackable, our method is restricted to a certain class of optimizations (Section 4.2): the optimized protocol $\mathcal{A}^\Delta$ differs from $\mathcal{A}$ with the addition of subactions in which the existing state of $\mathcal{A}$ is not mutated. Then, we develop an algorithm for automated porting the optimization, and the derived protocol is guaranteed correctness (Section 4.3). Last, we analyze a bunch of Paxos invariants and show the feasible optimizations (Section 4.4).

|  |  | **Raft*** | **MultiPaxos** |
|---|---|---|---|
| variables | Constant | Quorums | Quorums |
|  | per server | currentTerm<br>isLeader<br>entries with index ≤ commitIndex | ballot<br>phase1succeeded<br>chosenSet |
|  | per instance | entry.index<br>entry.val<br>entry.bal | instance.id<br>instance.val<br>instance.bal |
|  | messages | requestVote<br>requestVoteOK<br>(im/ex) append<br>(im/ex) appendOK | prepare<br>prepareOK<br>accept<br>acceptOK |
| functions |  | RequestVote | Phase1a |
|  |  | RecieveVote | Phase1b |
|  |  | BecomeLeader | Phase1Succeed<br>Phase2a<br>Phase2b |
|  |  | AppendEntries | Phase2a<br>Phase2b |
|  |  | RecieveAppend | Phase2b |
|  |  | LeaderLearn | Learn |

Fig. 3. Mapping between Raft* and MultiPaxos. "im" stands for implicit. "ex" stands for explicit

### 4.1 Problem definition

To automatically port some optimization across protocols, we must be able to describe protocols and optimization in a formal way. Only then can we define the problem of automated optimization porting.

**Specifying a protocol.** We specify a protocol as a state machine, which can be defined by its initial state and a set of allowed state transitions. We use the TLA$^+$ language [41] for specifying state machines in this paper. Any other state-based specification methods [12, 55] would also work.

In TLA$^+$, one specifies a protocol by describing the set of allowed state transitions called the *Next* action, represented as a collection of subactions $a_1 \vee a_2 \vee \ldots$ where $\vee$ is the or (disjunction) operator. Each subaction $a_i$ is a formula in conjunctive form with one or more clauses; the clauses specify the state transition's enabling conditions and the next state value. As an example, we consider a key-value store supporting two operations Put(k, v) and Get(k). Figure 4a shows its TLA$^+$ specification. The key-value store's internal state is a hash table (table) where each entry corresponds to a set. Its next-state action (Next) consists of two subactions Put(k,v) and Get(k), for each potential key and value. In Figure 4a, the subaction Put(k, v) is defined to equal ( $\triangleq$ ) the boolean formula asserting that the hash table entry for key $k$ must contain value $v$ in the next state (line 2). In TLA$^+$, attaching the ′ symbol with a variable represents its value in the new state. Hence the formula table'[k]={v} is true only when the hash table entry for key k in the next state equals to v. We use the output variable to represent value returned to users, thus Get(k) uses the clause output' = table[k] to assert the value of the output variable in the new state. In this example $\mathcal{A}$, there is no subaction involving more than one clause in the conjunctive form because there is no enabling conditions for either Put or Get. We will see more sophisticated subactions in subsequent examples.

| | |
|---|---|
| 1: VARIABLES table, output<br>2: Put(k, v) $\triangleq$ table'[k] = {v}<br>3: Get(k) $\triangleq$ output' = table[k]<br>4: Init $\triangleq$ ∀ k ∈ Nat: table[k] = {}<br>5: Next $\triangleq$ ∃ k ∈ Nat, v ∈ Values: Put(k, v) ∨ Get(k) | 1: VARIABLES logs, output<br>2: Write(i, v) $\triangleq$ (i = 0 ∨ logs[i-1] ≠ {})<br>3:           ∧ logs'[i] = v<br>4: Read(i) $\triangleq$ output' = logs[i]<br>5: Init $\triangleq$ ∀ i ∈ Nat : logs[i] = {}<br>6: Next $\triangleq$ ∃ i ∈ Nat, v ∈ Values: Write(i, v) ∨ Read(i) |
| (a) A key-value store ($\mathcal{A}$). | (b) The protocol $\mathcal{B}$ that stores data in a log. $\mathcal{B}$ refines the key-value store $\mathcal{A}$. |
| 1: VARIABLES table, output, size<br>2: Put(k, v) $\triangleq$ table[k] = {}<br>3:           ∧ table'[k] = {v}<br>4:           ∧ size' = size + 1<br>5: Get(k) $\triangleq$ output' = table[k]<br>6: Init $\triangleq$ ∀ k ∈ Nat: table[k] = {} ∧ size = 0<br>7: Next $\triangleq$ ∃ k ∈ Nat, v ∈ Values: Put(k, v) ∨ Get(k) | 1: VARIABLES logs, output, size<br>2: Write(i, v) $\triangleq$ logs[i] = {}<br>3:           ∧ (i = 0 ∨ logs[i-1] ≠ {})<br>4:           ∧ logs'[i] = {v}<br>5:           ∧ size' = size + 1<br>6: Read(i) $\triangleq$ output' = logs[i]<br>7: Init $\triangleq$ ∀ k ∈ Nat: logs[i] = {} ∧ size = 0<br>8: Next $\triangleq$ ∃ i ∈ Nat, v ∈ Values: Write(i, v) ∨ Read(i) |
| (c) The optimized protocol $\mathcal{A}^\Delta$. | (d) The generated TLA$^+$ specification of $\mathcal{B}^\Delta$. |

Fig. 4. The TLA$^+$ specifications of the example.

**Defining protocol equivalence.** With the specification, we are able to describe the refinement mapping formally. We use $\mathcal{B} \Rightarrow \mathcal{A}$ to refer that $\mathcal{B}$ has a refinement mapping to $\mathcal{A}$. (e.g., Raft* $\Rightarrow$ Paxos). In order for $\mathcal{B}$ to refine $\mathcal{A}$, we must show that each state in $\mathcal{B}$'s state space can be mapped to some state in $\mathcal{A}$'s state space such that any state transition sequence allowed by $\mathcal{B}$ corresponds to a valid state transition sequence in $\mathcal{A}$ under the mapping. More concretely, let $Var_\mathcal{A}$ (or $Var_\mathcal{B}$) represent the state variables of $\mathcal{A}$ (or $\mathcal{B}$). Let $f$ be some function that maps $\mathcal{B}$'s state space to $\mathcal{A}$'s, i.e., $Var_\mathcal{A} = f(Var_\mathcal{B})$. Suppose $a_i$ is some subaction in $\mathcal{A}$, we use the term $\overline{a_i}$ to refer to the conjunctive formula when we substitute $Var_\mathcal{A}$ in $a_i$ with $Var_\mathcal{A} = f(Var_\mathcal{B})$. If $\mathcal{B}$ refines $\mathcal{A}$ under $f$, then every subaction $b_i$ in $\mathcal{B}$ implies some subaction $a_j$ in $\mathcal{A}$ or a no-op step[1], i.e. $b_i \Rightarrow \overline{a_j} \lor f(Var'_b) = f(Var_b)$, where $\Rightarrow$ is the boolean operator for implication.

Figure 4b shows an example protocol $\mathcal{B}$ which stores data in a log. Subaction Append(i, v) stores a value at the end of the log at index i. The conjunctive clause at line 2 ensures the invariant that values are stored in the log continuously. Subaction Read(i) reads the log entry i. Protocol $\mathcal{B}$ in Figure 4b refines protocol $\mathcal{A}$ in Figure 4a under the state mapping that maps the $i$-th entry of the log to the hash table entry with key $k = i$. The subaction Append in $\mathcal{B}$ implies Put $\mathcal{A}$, and Read implies Get. The formal proof of the refinement between Raft* and Paxos can be found in [9].

**The problem of porting optimization.** Informally, given two "equivalent" protocols, $\mathcal{A}$ and $\mathcal{B}$, $\mathcal{B} \Rightarrow \mathcal{A}$, as well as an optimized version of protocol $\mathcal{A}$, we would like to adapt the optimization also to improve protocol $\mathcal{B}$. More importantly, we require the adaption to follow an automated procedure that guarantees the correctness of the resulting protocol.

To state the task formally, we are given some protocol $\mathcal{A}$, its optimized version $\mathcal{A}^\Delta$, and another protocol $\mathcal{B}$ which refines $\mathcal{A}$, all specified in TLA$^+$. Furthermore, we assume that all three protocols $\mathcal{A}$, $\mathcal{A}^\Delta$, $\mathcal{B}$ have been proven correct. The problem of porting an optimization is to automatically derive the TLA$^+$ specification of protocol $\mathcal{B}^\Delta$ such that $\mathcal{B}^\Delta$ improves the performance of $\mathcal{B}$ and is guaranteed to be correct.

---
[1]The no-op step is commonly called a stuttering step [1, 46]

The problem as stated above is very general, and we did not make any assumptions on the types of correctness proofs given for the protocols, nor on any formal relationships between $\mathcal{A}$ and $\mathcal{A}^\Delta$. As a first step towards tackling this problem, we are going to devise a solution that applies to a restricted class of optimization. It is a central challenge to find a restricted set of optimizations so as to enable a working solution.

We note that the goal is to derive an optimized protocol by giving its TLA$^+$ specification. We do not aim to automatically generate the implementation of the optimization protocol, but instead, leave the implementation to the developers.

### 4.2 Non-mutating optimization

We consider protocol $\mathcal{A}^\Delta$, which an optimized version of protocol $\mathcal{A}$. The optimization applied to $\mathcal{A}$ can be defined by the difference between the specification of $\mathcal{A}^\Delta$ and $\mathcal{A}$. In particular, the state variables of $\mathcal{A}^\Delta$ include all state variables of $\mathcal{A}$ and may contain additional variables introduced by the optimization. Each subaction of $\mathcal{A}^\Delta$ can be of three forms:

- *An added subaction.* This is a subaction that has no relationships to existing subactions in $\mathcal{A}$.
- *An unchanged subaction.* This is a subaction that is identical to an existing subaction in $\mathcal{A}$.
- *A modified subaction.* This is a subaction derived from an existing subaction in $\mathcal{A}$ by adding extra conjunctive clauses.[2]

Our proposed method for porting an optimization works for a restricted class of optimizations, which we refer to as *non-mutating optimization*. For an optimization $\mathcal{A}^\Delta$ to be considered as non-mutating, we require that none of its added subactions and none of the added clauses in its modified subactions mutate the state variables of $\mathcal{A}$ ($Var_\mathcal{A}$). The subactions are free to mutate the new state variables ($Var_\Delta$) added by $\mathcal{A}^\Delta$.

Figure 4c shows protocol $\mathcal{A}^\Delta$, as an example of non-mutating optimization on $\mathcal{A}$. The optimized protocol $\mathcal{A}^\Delta$ adds a new state variable size that tracks how many values have been stored in the hash table. Comparing Figure 4c with Figure 4a, we can see that $\mathcal{A}^\Delta$ adds the new clause (line 4) to existing subaction Put and no completely new subactions. As the new clause does not modify $\mathcal{A}$'s state (table), $\mathcal{A}^\Delta$ is a non-mutating optimization.

Non-mutating optimizations not only let us port optimization from $\mathcal{A}$ to $\mathcal{B}$ using the method described in Section 4.3, it also has the important advantage that $\mathcal{A}^\Delta$ can be shown to refine $\mathcal{A}$ under the identity state mapping function that ignores the extra state. Therefore, non-mutating optimizations can always be guaranteed correctness. By contrast, state-mutating optimization may or may not have a refinement mapping to $\mathcal{A}$, and thus its correctness requires a separate proof.

### 4.3 How to port non-mutating optimization

We only consider the case of porting non-mutating optimizations. Additionally, if the optimization reads the parameters of protocol $\mathcal{A}$, our method also requires a *parameter mapping* from $\mathcal{B}$ to $\mathcal{A}$.

**Parameter mapping.** Let $P_\mathcal{A}$ and $P_\mathcal{B}$ be the parameter variables of $\mathcal{A}$ and $\mathcal{B}$, respectively. We say $\mathcal{B}$ has a parameter mapping to $\mathcal{A}$ if there exists a function $f_{arg}$ that maps the arguments of subactions in $\mathcal{B}$ to the arguments of subactions in $\mathcal{A}$, i.e. $P_\mathcal{A} = f_{args}(P_\mathcal{B})$.[3] The extra clauses added in a modified subaction in $\mathcal{A}^\Delta$ may use parameter variables.

---

[2]If the derivation deletes an existing conjunctive clause, then the resulting subaction of $\mathcal{A}^\Delta$ must be viewed as an added subaction instead of a modified one.

[3]To put it more formally, given parameter mapping $P_\mathcal{A} = f_{args}(P_\mathcal{B})$, we use $\overline{Next_\mathcal{A}}$ to refer to the formula after substituting state variables $Var_a$ with $f(Var_b)$ and parameter variables $P_\mathcal{A}$ with $f_{args}(\mathcal{B})$. $f_{args}$ is a valid parameter mapping if $Next_\mathcal{B} \Rightarrow \overline{Next_\mathcal{A}}$.

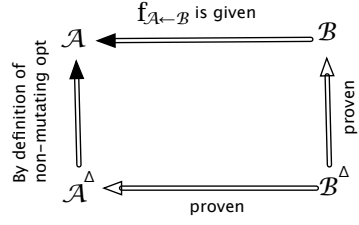

Fig. 5. The refinement mappings among given protocols. $\mathcal{A}^\Delta$ is an optimized version of $\mathcal{A}$ using non-mutating optimization. $\mathcal{B}^\Delta$ is the optimized version of $\mathcal{B}$ generated by our method in Section 4.3.

Therefore, the parameter mapping is required in order to correctly translate those clauses to be used in a corresponding subaction in $\mathcal{B}^\Delta$.

**Porting the optimization.** We are now ready to describe how to transform the specification of $\mathcal{A}^\Delta$ to create $\mathcal{B}^\Delta$, thereby porting the optimization from $\mathcal{A}$ to $\mathcal{B}$. First, we obtain $\mathcal{B}^\Delta$'s state variables as $Var_{\mathcal{B}^\Delta} = Var_\mathcal{B} \cup Var_\Delta$. We also obtain $\mathcal{B}^\Delta$'s initial state ($Init_{\mathcal{B}^\Delta}$) from $Init_\mathcal{B}$ and $Init_{\mathcal{A}^\Delta}$ by replacing every $\mathcal{A}$'s state variable using the state mapping $Var_\mathcal{A} = f(Var_\mathcal{B})$. Next, we generate the subactions of $\mathcal{B}^\Delta$ from each subaction $a_i^\Delta$ of $\mathcal{A}^\Delta$ and the no-op step. There are three cases:

*Case-1*: $a_i^\Delta$ is an added subaction. We turn $a_i^\Delta$ into a corresponding added subaction $b_i^\Delta$ by substituting state variable $Var_a$ with $f(Var_b)$ and keeping $Var_\Delta$ unchanged.

*Case-2*: $a_i^\Delta$ is an unchanged subaction equal to $a^i$ in $\mathcal{A}$, or the no-op step. There is a set of subactions in $\mathcal{B}$ that imply $a^i$ according to the refinement mapping $f_{\mathcal{B} \to \mathcal{A}}$. We add the set of subactions to $\mathcal{B}^\Delta$.

*Case-3*: $a_i^\Delta$ is a modified subaction of $a^i$ in $\mathcal{A}$. Again, there is a set of subactions in $\mathcal{B}$ that imply $a^i$ according to $f_{\mathcal{B} \to \mathcal{A}}$. Suppose $b_j$ is a subaction in the set. We add $b_j$ to $\mathcal{B}^\Delta$ if $b_j$ is not already added (in Case-2). Furthermore, we include the extra clauses added by $a_i^\Delta$ in $b_j$ by substituting $Var_a = f_{\mathcal{B} \to \mathcal{A}}(Var_b)$ and $P_a = f(P_b)$.

**Correctness.** We prove that our method generates a correct specification of the optimized protocol $\mathcal{B}^\Delta$. The proof contains two parts. First, we need to show that $\mathcal{B}^\Delta$ correctly incorporates the optimization in $\mathcal{A}^\Delta$. This can be proven by demonstrating that $\mathcal{B}^\Delta$ refines $\mathcal{A}^\Delta$, thus $\mathcal{B}^\Delta$ preserves the invariants introduced by the optimization. Second, we also need to show that $\mathcal{B}^\Delta$ remains correct w.r.t. the original protocol $\mathcal{B}$. This can be proven by demonstrating that $\mathcal{B}^\Delta$ refines $\mathcal{B}$, thus $\mathcal{B}^\Delta$ preserves the invariants of the original protocol $\mathcal{B}$.

As a summary, Figure 5 illustrates the refinement mappings that exist among the four protocols, $\mathcal{A}$, $\mathcal{B}$, $\mathcal{A}^\Delta$, $\mathcal{B}^\Delta$. Next, we provide proof sketches for the refinement mappings of $\mathcal{B}^\Delta$.

First, we proving $\mathcal{B}^\Delta$ refines $\mathcal{A}^\Delta$. $\mathcal{B}^\Delta$ refines $\mathcal{A}^\Delta$ under the state mapping function, which maps the state variables of $\mathcal{B}$ to those of $\mathcal{A}$ according to $f_{\mathcal{A} \leftarrow \mathcal{B}}$ and leaves the variables introduced by optimization $\Delta$ unchanged. To prove the correctness of this refinement mapping, we must show that $\mathcal{B}^\Delta$'s initial state implies $Init_{\mathcal{A}^\Delta}$, and $\mathcal{B}^\Delta$'s next-state action ($Next_{\mathcal{B}^\Delta}$) implies $Next_{\mathcal{A}^\Delta}$. The former implication is relatively straightforward, so we focus the discussion on the latter.

To show $Next_{\mathcal{B}^\Delta}$ implies $Next_{\mathcal{A}^\Delta}$, we show that each $\mathcal{B}^\Delta$'s subaction ($b_i^\Delta$) implies some $\mathcal{A}^\Delta$'s subaction or a no-op step. According to our method, $b_i^\Delta$ can be added to $\mathcal{B}^\Delta$ in one of three cases. For case-1 and 2, it is easy to show that $b_i^\Delta$ implies $a_i^\Delta$ or the no-op step by construction. In case-3, $b_i^\Delta$ is constructed from $b_i$ and a subaction $a_j^\Delta$ in $\mathcal{A}^\Delta$, such that $a_j^\Delta = a_j \wedge \Delta_{a_j}, b_i^\Delta = b_i \wedge \overline{\Delta_{a_j}}$. $\Delta_{a_j}$ is defined as the extra conjunctive clauses the optimization has added to $a_j$ to form $a_j^\Delta$. $\overline{\Delta_{a_j}}$ is obtained from $\Delta_{a_j}$ by substituting variables $Var_a = f_{\mathcal{B} \to \mathcal{A}}(Var_b)$ and parameters $P_a$ by $f_{args}(P_b)$. Because of $b_i \Rightarrow \overline{a_j}$, we have $b_i^\Delta \Rightarrow \overline{a_j} \wedge \overline{\Delta_{a_j}}$ which is equivalent to $b_i^\Delta \Rightarrow \overline{a_j^\Delta}$.

Then, we prove $\mathcal{B}^\Delta$ refines $\mathcal{B}$. We argue that optimized protocol $\mathcal{B}^\Delta$ refines $\mathcal{B}$ because the optimizations of $\mathcal{B}^\Delta$ over $\mathcal{B}$ are non-mutating optimizations. Thus, the state mapping function that simply drops the new variables added by the optimization results in a valid refinement mapping. Why $\mathcal{B}^\Delta$ is a non-mutating optimization can be shown by analyzing the three cases of our method. We omit the details here.

### 4.4 Paxos variants and optimizations.

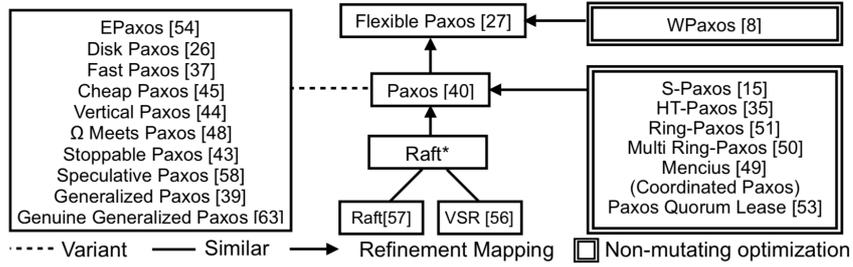

Fig. 6. The relationship of different consensus protocols, and non-mutating optimizations on Paxos.

Let us now examine the landscape of existing Paxos variants and optimizations using the lens of our method. We studied all known Paxos variants and optimizations, and Figure 6 shows the relationship between these protocols. In the center of Figure 6, we have the canonical Paxos protocol.

Among the protocols we studied, 6 protocols belong to the class of non-mutating optimization on Paxos. They are shown in the double-lined box in Figure 6 and are the candidates for our porting method. Flexible Paxos [27] relaxes the majority quorum restriction in Paxos to allow differently sized quorums as long as the quorums used in the two phases of Paxos exchange are guaranteed to intersect. As a result, Paxos refines Flexible Paxos but not the other way around. WPaxos is a recently proposed non-mutating optimization on Flexible Paxos. Therefore, our method would allow us to port the optimization of WPaxos to Paxos.

As for the rest of the protocol variants (shown in the left-most box in Figure 6), their relationships to Paxos cannot be captured by refinement mapping. The reasons for the lack of refinement mapping are varied. For example, Fast Paxos changes the quorum size of Paxos to include a super-majority, which prevents a refinement mapping from Fast Paxos to Paxos. However, it also misses state transitions allowed in Paxos, which precludes a refinement mapping from Paxos to Fast Paxos.

Next, we choose two optimizations as case studies: Paxos Quorum Lease and Mencius. We explain what these opimizations are and how to port them to Raft. In following discussion, we adopt the pseudocode instead of TLA$^+$ syntax for simplicity. A complete discussion (including the refinement mapping, pseudocode, and TLA$^+$) can be found in the long version [9].

**Paxos Quorum Lease.** In Paxos, a strongly consistent read operation is performed by persisting the operation into the log as if it were a write. Paxos Quorum Lease (PQL) [53] introduces an optimization that allows any replica to read locally if the replica holds leases from a quorum of replicas (*quorum-lease*).

Quorum-lease can co-exist perfectly with the quorum in Paxos. Any replica can grant a lease. A replica considers itself holding a quorum-lease if it holds leases from a quorum of replicas. Any lease-quorum must overlap with any Paxos quorum (usually both quorums are majorities of replicas). In Paxos any commit needs to collect from a quorum of acknowledgments, which will intersect with the lease quorum. Therefore, as long as we require every replica in a Paxos quorum to notify their granted lease holders before the replica commits any values, the system is safe—both read and write are consistent.

PQL is a non-mutating variant of Paxos, because all its added and modified subactions are not changing the state variables in Paxos. Figure 11 shows the algorithm changes introduced by PQL. The actions that are changed is Phase2b and Learn, where extra checks on the lease quorum are performed. The added actions are Read and LocalRead, which are wrappers for the client and server to perform read operations at the local replica.

```
1   function Read(c, k):
2     c sends <''localRead'', k> to 1 server s
3     if c receives <''ReadReply'', v> from s
4     then
5       return v
6
7   function LocalRead(s):
8     if s receives <''localRead'', k>
9        && validLeasesNum ≥ f + 1
10       &&all instances modifed k are in chosenSet
11    then
12       s replies <''ReadReply'', LocalCopy(k)>
13
14  function Phase2b(s):
15       ...
16       s replies <''acceptOk'',..., leases granted by s>
17
18  function Learn(s):
19    if s receives <''acceptOk'', i, v, b, s, leases> from f + 1 acceptors
20    then
21       holderSet = holders of received leases
22
23    if s receives <''acceptOk'', ... > from all holders in holderSet
24    then
25       ...
```

Fig. 7. Paxos Quorum Lease

**Raft*-PQL.** Figure 13 shows the algorithm after applying PQL to Raft*. The code in blue shows the changed part after porting the code to Raft*. For a replica to perform a local read, the replica needs to check if two conditions hold. First, the replica must be holding leases from at least $f + 1$ replicas (including itself). Second, the replica needs to wait until commitIndex is greater than the largest index of entries which modify the target record (line 4 in Figure 13). This is transformed from PQL where all modifications must be in the chosenSet (line 10 in Figure 11).

A replica attaches the lease holders granted by itself in appendOk message, which maps to the acceptOk message. In LeaderLearn, the leader needs to collect the holders of leases attached in the messages and granted by itself (line 13, 14). This is because the $f$ appendOk messages with one extra implicit appendOk message imply $f + 1$ acceptOk message in Paxos. Thus, collecting leases attached in $f + 1$ messages (line 21 in Figure 11) should be transformed into collecting the leases from $f$ messages and local granted (the implicit message).

```
1  function LocalRead(s):
2    if s receives <‘‘localRead’’, k>
3       && s.validLeasesNum ≥ f + 1
4       && indexes of entries in s.log modified k ≤ s.commitIndex
5    then
6       ...
7
8  function LeaderLearn(s):
9    if s receives <‘‘appendOK’’, t, index, holders> from f acceptors
10      && s.isLeader
11      && s.currentTerm == t
12   then
13      holderSet = received holders ∪ holders of leases granted by the leader
14   if s receives <‘‘appendOK’’, ... > from all holders in holderSet
15   then
16      ...
```

Fig. 8. Raft* Quorum Lease

Before the automated version of Raft*-PQL, we had a handworked version of applying PQL to Raft. By comparing these two versions, we find that the handworked version has a few subtle errors. For example, in Raft*, a leader only waits for responses from $f$ replicas for its AppendEntry requests, which does not include itself. Our handworked version just uses the granted information from these $f$ replicas, and ignore the lease holders granted by the leader. However, with our algorithm, as $f + 1$ accept messages are mapped to $f$ append messages and the receiver is the leader. Thus we are able to include lease holder granted by the leader automatically.

**Mencius.** Multi-Paxos requires all clients requests to be sent to a leader for better throughput. This could lead to unbalanced load between the leader replica and other replicas. When replicas are located in different data centers, non-leader replicas will need at least two wide-area round-trips to commit any requests because requests need to be forwarded to the leader. To address these issues, Mencius [49] partitions the Paxos instances so that each replica serves as the default leader for a distinct subset of instances. With geo-replicas, a client can send its requests to the nearest replica. The replica can commit these requests using those Paxos instances for which it is the default leader. Thus, Mencius can balance the load among all replicas and also reduces wide-area round-trips.

Mencius partitions the instance (log) space in a round-robin way. For example, with three replicas $r_1, r_2, r_3$, $r_1$ is the default leader for log entries $(0, 3, 6, ...)$, $r_2$ is the leader for $(1, 4, 7, ...)$, and $r_3$ for $(2, 5, 8, ...)$. Mencius separates the log entry execution from its commit. The log is still executed sequentially and each replica keeps committing skip to keep the system moving forward. To prevent a crashed replica from delaying the system, the instances belong to one replica can be committed no-op by other replicas. These optimizations can help Mencius to commit and execute requests within 1.5 round-trips on average. Due to space limitation here, the pseudo-code of Mencius with highlighted added/modified subactions, as well as more details about why it is a non-mutating variant, can be found in the attached supplemental material.

**Raft*-Mencius.** The complete version of Raft*-Mencius and other similar optimizations are included in [9]. We only describe some interesting details here. In addition to the Paxos state variables, each replica needs to keep an array of "skip-tags", which indicates that which log entries can be skipped. When a replica becomes the leader, it needs to collect not only values but also skip-tags from other replicas. Because Phase2b action in Paxos corresponds to many actions ( AppendEntries, ReceiveAppend) in Raft*, whatever changes Mencius makes to Phase2b should be applied to these actions as well. As an example, if the new appended entries are nop and from default leader, these entries should be marked executable.

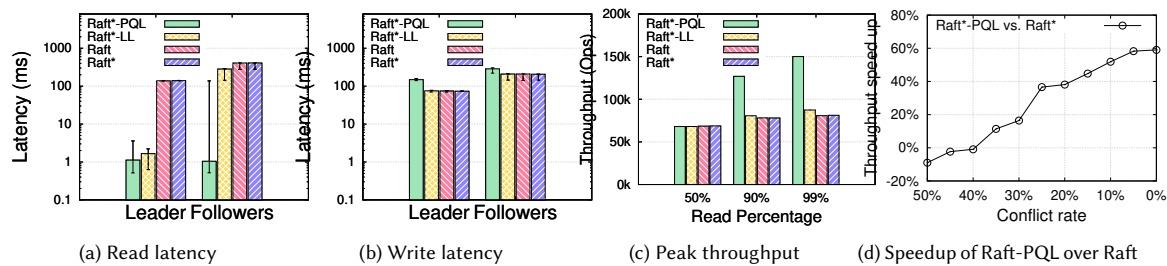

(a) Read latency  (b) Write latency  (c) Peak throughput  (d) Speedup of Raft-PQL over Raft

Fig. 9. Raft-PQL vs. LL vs. Raft. Each bar in (a) and (b) represents the 90th percentile latency of the requests with an error bar from the 50th to 99th percentiles. The y-axis is in log scale for (a) and (b).

Because Phase2b in Paxos is implied by multiple sub-actions in Raft*, it is possible for a handworked solution in porting Mencius to miss some of the actions. For example, if the handworked solution only applies changes on Phase2b in Paxos to ReceiveAppend in Raft* (missing AppendEntries), the solution could miss some optimization opportunities or even generate an incorrect protocol.

## 5 EVALUATION

This section shows that the generated algorithms achieve similar optimization effects with their Paxos counterparts [53, 49].

**Testbed.** The experiments were conducted on Amazon EC2 across 5 different geographical regions: Oregon, Ohio, Ireland, Canada, Seoul. In each region, two m4.xlarge instances are used for the client and server processes respectively. Each instance has 4 virtual CPUs, 16GB memory and one SSD with 750 Mbps bandwidth. The latency across sites varies from 25ms to 292ms.

**Workload.** Our evaluation uses closed-loop clients with a YCSB [22] alike workload: each client issues get or put requests back-to-back. The system is initialized with 100K records. To simulate contention, each client accesses the same popular record at a configured rate (i.e., conflict rate). When not accessing the popular record, the key space is pre-partitioned among the datacenters evenly, and a key is selected from this partition with uniform probability. Raft*-PQL is evaluated with 90% read by default. For Raft*-Mencius, we use a workload with 100% writes. Each trial is run for 50 seconds with 10 seconds for both warm-up and cool-down. Each number reported is the median in 5 trials.

**Implementation.** The implementation of Raft* is based on a popular industrial Raft codebase—etcd (version c4fc8c09). etcd has a few important optimizations. First, it has a customized network layer for efficient communication. Second, when a follower receives multiple requests from clients, it forwards them to the leader in a batch. Such techniques improve the system throughput when follower servers accept client requests. In our tests, etcd is 2.4× better in throughput when these optimizations are turned on. We keep these optimizations on in our tests to give etcd extra advantages. Oregon is used as the leader site for etcd which gives it the best result since Oregon has the best network condition.

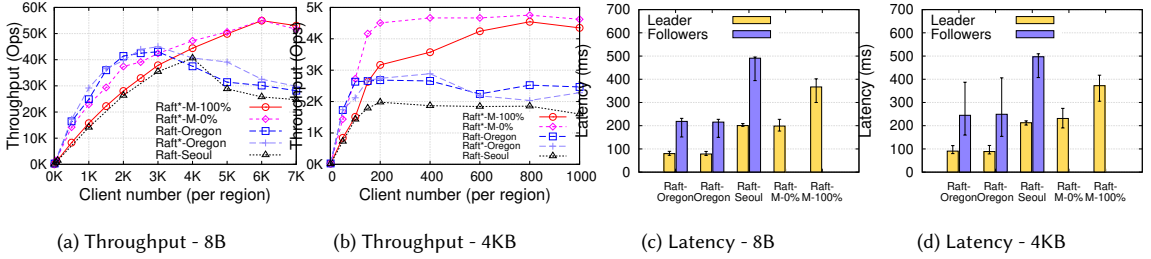

(a) Throughput - 8B  (b) Throughput - 4KB  (c) Latency - 8B  (d) Latency - 4KB

Fig. 10. Raft-M vs. Raft. Raft-M-100% and Raft-M-0% stand for Raft-Mencius with workload under 100% and 0% conflict rate. Raft-Oregon and Raft-Seoul stand for the leader is in Oregon or Seoul. 8B and 4KB are the request size of the workload.

### 5.1 Raft*-PQL

We evaluate Raft*-PQL with the same lease parameters in [53]: the lease duration is 2 seconds, and the grantors renew their leases every 0.5 seconds. In addition to Raft and Raft*, we also compare Raft*-PQL with Leader Lease (LL). Here the leader has sole ownership of the lease, so only the leader can process a read request with its local copy. We use 90% read workload with a 5% conflict rate by default.

**Latency.** First, we compare the latency with 50 clients per region. In Raft*-PQL, any server with an active lease is able to conduct local consistent reads, thus 90% of the read requests have only 1ms latency (Figure 9a). In comparison, for LL, only the leader can process read request with similar latency (1.6 ms). Raft*-PQL has 1% read requests on the follower with high latency (∼137ms). This is caused by the 5% contention in the workload: for Raft*-PQL, followers need to wait for all concurrent conflicting write requests to be applied before processing the read request. Raft* has similar latency with Raft, as they use 1RT to process a read request. For write latency (Figure 9b) Raft*-PQL is a little bit higher than others, as it needs to wait for leaseholders' acknowledge to commit a write operation, while others can always choose the fastest majority.

**Throughput.** Figure 9c shows how is the peak throughput affected by reading percentage(50%, 90%, and 99%). Raft, Raft* and LL achieve almost the same peak throughput, as the leader's CPU is the bottleneck, and the saturated leader CPU has the same capability to handle read and write requests for these algorithms. In contrast, Raft*-PQL achieves 1.6× and 1.9× speedup with 90% and 99% reads. The throughput advantage of Raft*-PQL is due to the fact that leader and followers are able to conduct read requests locally. Figure 9d also shows how is the throughput speedup affected by the conflict rate. The figure does not show the speedup of Raft*-PQL over Raft*, as they are similar. The speedup increases with the decreasing of conflict rate since all followers can return read requests to the user immediately instead of wait for the commit of the conflicting write.

### 5.2 Raft*-Mencius

Similar with Mencius, Raft*-Mencius also supports the optimization for commutative operations. Thus, we do the experiment under different contention levels.

**Throughput.** We use a 100% put workload to measure Raft*-Mencius with 0% and 100% contention, marked as Raft*-Mencius-0% and Raft*-Mencius-100% respectively. To make a fair comparison, we evaluate both the best and worst case scenarios for Raft in the wide area by placing the leader in the nearest (Oregon) and farthest (Seoul) servers to all other regions (Raft-Oregon and Raft-Seoul). We only evaluate Raft* with the leader at Oregon for reference.

Figure 10a gives the throughput when the system is bounded by the CPU. Raft*-Mencius can reach a peak throughput of 55K operations per second (ops) since it balances the load among all replicas. In contrast, other systems can reach the peak throughput of 41K ops after their leaders' CPUs are saturated. Figure 10b gives the throughput when the system is network bounded. Raft reaches the peak throughput after saturating leader's network bandwidth. Raft-Oregon has 30% higher throughput than Raft-Seoul as Oregon has higher bandwidth. Raft*-Mencius has 70% higher throughput than Raft-Oregon because it is able to utilize all replicas' network bandwidth. In both figures, with a small number of clients, Raft-Oregon and Raft*-Mencius-0% have better performance than others due to their lower latency.

**Latency.** Figure 10c and Figure 10d show the latency with 50 clients per region. Among all configurations, the leader of Raft-Oregon processes requests with the lowest latency (79ms), as the quorum of Oregon, Ohio and Canada are closest to each other. In comparison, Raft*-Mencius-100% has much higher 90% percentile latency. This is because, to commit an entry, a server needs to learn all other servers' commit decisions on previous entries. Raft*-Mencius-0% has lower latency since it only needs to wait for other servers to send append or skip requests. However, its latency is still bounded by the farthest server in Seoul (360ms vs. 110ms).

## 6 RELATED WORK

**Elementary consensus protocols.** In addition to Raft [57], Paxos [40, 42] has been the de facto standard for implementing state machine replication [60, 11, 23, 64, 16, 67, 34, 59]. However, Paxos is not the only solution. There are many alternative protocols. For example, Viewstamped Replication (VR) [56] was published earlier than Paxos, and ZooKeeper [30] uses ZAB [31]. Canonical Paxos is presented in a very abstract style; it ignores many of the engineering details. This has led to many efforts in filling in the blanks for real system designs. Renesse et al. [65] attempt to concretely express the algorithmic details of the MultiPaxos protocol. MaziÃİres [52], Kirsch [32], and Chandra [20] expand on some of the practical aspects of building a distributed system built with Paxos. Raft makes understandability a primary concern [57], and a clean-room implementation and evaluation of Raft are provided [28]. Moreover, Afek et al. [6] decompose the consensus algorithm into a common framework with simple blocks.

**Paxos variants and optimizations.** Figure 6 has shown a number of Paxos variants. Among the non-mutating variants, WPaxos [8] partitions object and uses flexible quorums for geo-replication [27]; HT-Paxos [35] and S-Paxos [15] assigns ordering tasks to multiple servers to remove bottlenecks. Ring Paxos [51] and Multi Ring-Paxos [50] partition the workload and achieve better performance. Among the mutating Paxos variants: Cheap Paxos [45] introduces auxiliary servers. Ω meets Paxos [48] elects a stable leader in a weak network environment. NetPaxos [25] adapts Paxos to SDN. Stoppable Paxos [43] is able to perform reconfiguration without slowing down. Additionally, Shraer et al. [61] and Vertical Paxos [44] discusses how to reconfigure a replicated state machine. Disk Paxos [26] achieves consensus in a disk cluster. Fast Paxos [37] and Multi-coordinated Paxos [17] introduce a fast quorum to reach consensus with a single round-trip. Generalized Paxos [39], Genuine Generalized Paxos [63] and EPaxos [54] resolve conflicts because execution. Speculative Paxos [58] introduces speculative execution when messages are delivered in order.

**Algorithm clarification and comparison.** Renesse et al. [66] compared Paxos to VR and ZAB using *refinement mapping*. Howell et al. [29] present proof of Paxos with replica-set-specific views. Castro et al. [18] proved the correctness of PBFT in a formalized method. The equivalence between Byzantine Paxos and PBFT [19] is discussed in [36]. Song et al. [62] identified common traits in the Paxos, Chandra-Toueg [21], and Ben-Or [14, 7] consensus algorithms. Abraham and Malkhi [2, 3, 4] discussed the connections between BFT consensus protocols and block-chain protocols. Compared to these works, there are two notable differences in this work: we have used a formal method TLA$^{+}$ [41] to model the refinement mappings [1] and provided a machine-checkable proof using TLAPS [24]; we have mechanically exported the optimizations from one family of protocols to another.

## 7 CONCLUSION

In this paper, we formalize the connection between Raft and Paxos. With the correspondences, we presented an automated approach to port optimizations from one consensus protocol to another. As case studies, we porting two optimizations from Paxos (PQL and Mencius) to Raft.

## REFERENCES

[1] M. Abadi and L. Lamport. "The existence of refinement mappings". In: *Theoretical Computer Science* 82.2 (1991).

[2] I. Abraham and D. Malkhi. "BVP: Byzantine Vertical Paxos". In: *Distributed Cryptocurrencies and Consensus Ledgers (DCCL)* (2016).

[3] I. Abraham, D. Malkhi, et al. "The blockchain consensus layer and BFT". In: *Bulletin of EATCS* 3.123 (2017).

[4] I. Abraham, D. Malkhi, K. Nayak, L. Ren, and A. Spiegelman. "Solida: A cryptocurrency based on reconfigurable byzantine consensus". In: *Proc. OPODIS*. 2017.

[5] Abraham, Chockler, Keidar, and Malkhi. "Byzantine Disk Paxos: Optimal Resilience with Byzantine Shared Memory". In: *Distributed Computing* (2006).

[6] Y. Afek, J. Aspnes, E. Cohen, and D. Vainstein. "Brief announcement: object oriented consensus". In: 2017.

[7] M. K. Aguilera and S. Toueg. "The correctness proof of Ben-Or's randomized consensus algorithm". In: 25.5 (2012).

[8] A. Ailijiang, A. Charapko, M. Demirbas, and T. Kosar. "WPaxos: Ruling the Archipelago with Fast Consensus". In: (2017).

[9] *Appendix*. https://github.com/raftpaxos/raftpaxos/blob/master/appendix.pdf.

[10] V. Arora, T. Mittal, D. Agrawal, A. El Abbadi, X. Xue, et al. "Leader or Majority: Why have one when you can have both? Improving Read Scalability in Raft-like consensus protocols". In: *9th {USENIX} Workshop on Hot Topics in Cloud Computing (HotCloud 17)*. 2017.

[11] J. Baker, C. Bond, J. Corbett, J. Furman, A. Khorlin, J. Larson, J.-M. Léon, Y. Li, A. Lloyd, and V. Yushprakh. "Megastore: providing scalable, highly available storage for interactive services". In: 2011.

[12] B. Barras, S. Boutin, C. Cornes, J. Courant, J.-C. Filliatre, E. Gimenez, H. Herbelin, G. Huet, C. Munoz, C. Murthy, et al. "The Coq proof assistant reference manual: Version 6.1". PhD thesis. Inria, 1997.

[13] R. Batra. "Implementation and evaluation of Paxos and Raft distributed consensus protocols". PhD thesis. 2017.

[14] M. Ben-Or. "Another advantage of free choice (extended abstract): completely asynchronous agreement protocols". In: 1983.

[15] M. Biely, Z. Milosevic, N. Santos, and A. Schiper. "S-Paxos: offloading the leader for high throughput state machine replication". In: *SRDS*. 2012.

[16] W. J. Bolosky, D. Bradshaw, R. B. Haagens, N. P. Kusters, and P. Li. "Paxos replicated state machines as the basis of a high-performance data store". In: 2011.

[17] L. J. Camargos, R. M. Schmidt, and F. Pedone. "Multicoordinated Paxos". In: 2007.

[18] M. Castro and B. Liskov. *A Correctness proof for a practical byzantine-fault-tolerant replication algorithm*. Tech. rep. 1999.

[19] M. Castro and B. Liskov. "Practical byzantine fault tolerance". In: 1999.

[20] T. D. Chandra, R. Griesemer, and J. Redstone. "Paxos made live: an engineering perspective". In: 2007.

[21] T. D. Chandra and S. Toueg. "Unreliable failure detectors for reliable distributed systems". In: 43.2 (1996).

[22] B. F. Cooper, A. Silberstein, E. Tam, R. Ramakrishnan, and R. Sears. "Benchmarking cloud serving systems with YCSB". In: 2010.

[23] J. C. Corbett et al. "Spanner: Google's globally distributed database". In: 2012.

[24] D. Cousineau, D. Doligez, L. Lamport, S. Merz, D. Ricketts, and H. Vanzetto. "TLA+ proofs". In: *International Symposium on Formal Methods*. 2012.

[25] H. T. Dang, D. Sciascia, M. Canini, F. Pedone, and R. Soulé. "Netpaxos: Consensus at network speed". In: *Proceedings of the 1st ACM SIGCOMM Symposium on Software Defined Networking Research*. 2015.

[26] E. Gafni and L. Lamport. "Disk paxos". In: (2003).

[27] H. Howard, D. Malkhi, and A. Spiegelman. "Flexible Paxos: Quorum intersection revisited". In: *CoRR* abs/1608.06696 (2016). URL: http://arxiv.org/abs/1608.06696.

[28] H. Howard, M. Schwarzkopf, A. Madhavapeddy, and J. Crowcroft. "Raft refloated: do we have Consensus?" In: 49.1 (2015).

[29] J. Howell, J. Lorch, and J. J. Douceur. "Correctness of Paxos with replica-set-specific views". In: (2004).

[30] P. Hunt, M. Konar, F. P. Junqueira, and B. Reed. "ZooKeeper: wait-free coordination for internet-scale systems". In: 2010.

[31] F. P. Junqueira, B. C. Reed, and M. Serafini. "Zab: high-performance broadcast for primary-backup systems". In: 2011.

[32] J. Kirsch and Y. Amir. "Paxos for system builders: an overview". In: *LADIS*. 2008.

[33] N. Koufos, A. Matakos, and T. Pappas. "Raft & RethinkDB: A thorough Examination". In: (2017).

[34] T. Kraska, G. Pang, M. J. Franklin, S. Madden, and A. Fekete. "MDCC: multi-data center consistency". In: 2013.

[35] V. Kumar and A. Agarwal. "HT-Paxos: high throughput state-machine replication protocol for large clustered data centers". In: *The Scientific World Journal* (2015).

[36] L. Lamport. "Byzantizing Paxos by refinement". In: 2011.

[37] L. Lamport. *Fast Paxos*. Tech. rep. MSR-TR-2005-112. Microsoft Research, 2005.

[38] L. Lamport. "Fast Paxos". In: 19.2 (2006).

[39] L. Lamport. *Generalized consensus and Paxos*. Tech. rep. MSR-TR-2005-33. Microsoft Research, 2005.

[40] L. Lamport. "Paxos made simple". In: 32.4 (2001), pp. 18–25.

[41] L. Lamport. *Specifying systems: the TLA+ language and tools for hardware and software engineers*. Addison-Wesley Longman Publishing Co., Inc., 2002.

[42] L. Lamport. "The part-time parliament". In: 16.2 (1998).

[43] L. Lamport, D. Malkhi, and L. Zhou. "Reconfiguring a state machine". In: *ACM SIGACT News* (2010).

[44] L. Lamport, D. Malkhi, and L. Zhou. "Vertical Paxos and primary-backup replication". In: 2009.

[45] L. Lamport and M. Massa. "Cheap Paxos". In: 2004.

[46] L. Lamport and S. Merz. "Auxiliary variables in TLA+". In: *arXiv preprint arXiv:1703.05121* (2017).

[47] J. Li, E. Michael, N. K. Sharma, A. Szekeres, and D. R. Ports. "Just say NO to Paxos overhead: Replacing consensus with network ordering". In: 2016.

[48] D. Malkhi, F. Oprea, and L. Zhou. "Ω meets paxos: Leader election and stability without eventual timely links". In: 2005.

[49] Y. Mao, F. P. Junqueira, and K. Marzullo. "Mencius: building efficient replicated state machines for WANs". In: 2008.

[50] P. J. Marandi, M. Primi, and F. Pedone. "Multi-ring paxos". In: 2012.

[51] P. J. Marandi, M. Primi, N. Schiper, and F. Pedone. "Ring Paxos: A high-throughput atomic broadcast protocol". In: 2010.

[52] D. Mazieres. *Paxos made practical*. Tech. rep. 2007.

[53] I. Moraru, D. G. Andersen, and M. Kaminsky. "Paxos quorum leases: Fast reads without sacrificing writes". In: 2014.

[54] I. Moraru, D. G. Andersen, and M. Kaminsky. "There is more consensus in egalitarian parliaments". In: *Proc. SOSP*. 2013.

[55] T. Nipkow, L. C. Paulson, and M. Wenzel. *Isabelle/HOL: a proof assistant for higher-order logic*. 2002.

[56] B. M. Oki and B. H. Liskov. "Viewstamped replication: A new primary copy method to support highly-available distributed systems". In: 1988.

[57] D. Ongaro and J. K. Ousterhout. "In search of an understandable consensus algorithm". In: 2014.

[58] D. R. Ports, J. Li, V. Liu, N. K. Sharma, and A. Krishnamurthy. "Designing distributed systems using approximate synchrony in data center networks". In: 2015.

[59] J. Rao, E. J. Shekita, and S. Tata. "Using Paxos to build a scalable, consistent, and highly available datastore". In: 4.4 (2011).

[60] F. B. Schneider. "Implementing fault-tolerant services using the state machine approach: a tutorial". In: 22.4 (1990).

[61] A. Shraer, B. Reed, D. Malkhi, and F. P. Junqueira. "Dynamic reconfiguration of primary/backup clusters". In: 2012.

[62] Y. J. Song, R. van Renesse, F. B. Schneider, and D. Dolev. "The building blocks of consensus". In: *IEEE ICDCN*. 2008.

[63] P. Sutra and M. Shapiro. "Fast genuine generalized consensus". In: *2011 IEEE 30th International Symposium on Reliable Distributed Systems*. IEEE. 2011, pp. 255–264.

[64] A. Thomson, T. Diamond, S.-C. Weng, K. Ren, P. Shao, and D. J. Abadi. "Calvin: fast distributed transactions for partitioned database systems". In: 2012.

[65] R. Van Renesse and D. Altinbuken. "Paxos made moderately complex". In: 47.3 (Feb. 2015).

[66] R. Van Renesse, N. Schiper, and F. B. Schneider. "Vive la différence: Paxos vs. viewstamped replication vs. zab". In: *IEEE Transactions on Dependable and Secure Computing* 12.4 (2015).

[67] I. Zhang, N. K. Sharma, A. Szekeres, and A. K. and Dan R. K. Ports. "Building consistent transactions with inconsistent replication". In: *Proc. SOSP*. 2015.

## A TWO OPTIMIZATIONS ON PAXOS AND HOW TO PORT THEM TO RAFT*

### A.1 Paxos Quorum Lease

In Paxos, a strongly consistent read operation is performed by persisting the operation into the log as if it were a write. Paxos Quorum Lease (PQL) introduces an optimization that allows any replica to read locally if the replica holds leases from a quorum of replicas (*quorum-lease*).

Quorum-lease can co-exist perfectly with the quorum in Paxos. Any replica can grant a lease. A replica considers itself holding a quorum-lease if it holds leases from a quorum of replicas. Any lease-quorum must overlap with any Paxos quorum (usually both quorums are majorities of replicas). In Paxos any commit needs to collect from a quorum of acknowledgments, which will intersect with the lease quorum. Therefore, as long as we require every replica in a Paxos quorum to notify their granted lease holders before the replica commits any values, the system is safe—both read and write are consistent.

PQL is a non-mutating variant of Paxos, because all its added and modified subactions are not changing the state variables in Paxos. Figure 11 shows the algorithm changes introduced by PQL. The actions that are changed is Phase2b and Learn, where extra checks on the lease quorum are performed. The added actions are Read and LocalRead, which are wrappers for the client and server to perform read operations at the local replica.

```
1   function Read(c, k):
2     c sends <''localRead'', k> to 1 server s
3     if c receives <''ReadReply'', v> from s
4     then
5       return v
6
7   function LocalRead(s):
8     if s receives <''localRead'', k>
9        && validLeasesNum ≥ f + 1
10       &&all instances modifed k are in chosenSet
11    then
12      s replies <''ReadReply'', LocalCopy(k)>
13
14  function Phase2b(s):
15      ...
16      s replies <''acceptOk'',..., leases granted by s>
17
18  function Learn(s):
19    if s receives <''acceptOk'', i, v, b, s, leases> from f + 1 acceptors
20    then
21      holderSet = holders of received leases
22
23    if s receives <''acceptOk'', ... > from all holders in holderSet
24    then
25      ...
```

Fig. 11. Paxos Quorum Lease

### A.2 Raft*-PQL

Figure 13 shows the algorithm after applying PQL to Raft*. The code in blue shows the changed part after porting the code to Raft*. For a replica to perform a local read, the replica needs to check if two conditions hold. First, the replica must be holding leases from at least $f + 1$ replicas (including itself). Second, the replica needs to wait until commitIndex

```
1  function LocalRead(s):
2    if s receives <‘‘localRead’’, k>
3       && s.validLeasesNum ≥ f + 1
4       && indexes of entries in s.log modified k ≤ s.commitIndex
5    then
6       ...
7
8  function LeaderLearn(s):
9    if s receives <‘‘appendOK’’, t, index, holders> from f acceptors
10      && s.isLeader
11      && s.currentTerm == t
12   then
13      holderSet = received holders ∪ holders of leases granted by the leader
14   if s receives <‘‘appendOK’’, ... > from all holders in holderSet
15   then
16      ...
```

Fig. 12. Raft* Quorum Lease

is greater than the largest index of entries which modify the target record (line 4 in Figure 13). This is transformed from PQL where all modifications must be in the chosenSet (line 10 in Figure 11).

A replica attaches the lease holders granted by itself in appendOk message, which maps to the acceptOk message. In LeaderLearn, the leader needs to collect the holders of leases attached in the messages and granted by itself (line 13, 14). This is because the $f$ appendOk messages with one extra implicit appendOk message imply $f + 1$ acceptOk message in Paxos. Thus, collecting leases attached in $f + 1$ messages (line 21 in Figure 11) should be transformed into collecting the leases from $f$ messages and local granted (the implicit message).

Before the automated version of Raft*-PQL, we had a handworked version of applying PQL to Raft. By comparing these two versions, we find that the handworked version has a few subtle errors. For example, in Raft*, a leader only waits for responses from $f$ replicas for its AppendEntry requests, which does not include itself. Our handworked version just uses the granted information from these $f$ replicas, and ignore the lease holders granted by the leader. However, with our algorithm, as $f + 1$ accept messages are mapped to $f$ append messages and the receiver is the leader. Thus we are able to include lease holder granted by the leader automatically.

```
1  function LocalRead(s):
2    if s receives <‘‘localRead’’, k>
3       && s.validLeasesNum ≥ f + 1
4       && indexes of entries in s.log modified k ≤ s.commitIndex
5    then
6       ...
7
8  function LeaderLearn(s):
9    if s receives <‘‘appendOK’’, t, index, holders> from f acceptors
10      && s.isLeader
11      && s.currentTerm == t
12   then
13      holderSet = received holders ∪ holders of leases granted by the leader
14   if s receives <‘‘appendOK’’, ... > from all holders in holderSet
15   then
16      ...
```

Fig. 13. Raft* Quorum Lease

### A.3 Mencius

Multi-Paxos requires all clients requests to be sent to a leader for better throughput. This could lead to unbalanced load between the leader replica and other replicas. When replicas are located in different data centers, non-leader replicas will need at least two wide-area round-trips to commit any requests because requests need to be forwarded to the leader. To address these issues, Mencius partitions the Paxos instances so that each replica serves as the default leader for a distinct subset of instances. With geo-replicas, a client can send its requests to the nearest replica. The replica can commit these requests using those Paxos instances for which it is the default leader. Thus, Mencius can balance the load among all replicas and also reduces wide-area round-trips.

Mencius partitions the instance (log) space in a round-robin way. For example, with three replicas $r_1, r_2, r_3$, $r_1$ is the default leader for log entries $(0, 3, 6, ...)$, $r_2$ is the leader for $(1, 4, 7, ...)$, and $r_3$ for $(2, 5, 8, ...)$. Mencius separates the log entry execution from its commit. The log is still executed sequentially and each replica keeps committing skip to keep the system moving forward. To prevent a crashed replica from delaying the system, the instances belong to one replica can be committed no-op by other replicas. These optimizations can help Mencius to commit and execute requests within 1.5 round-trips on average. Due to space limitation here, the pseudo-code of Mencius with highlighted added/modified subactions, as well as more details about why it is a non-mutating variant, can be found in the attached supplemental material.

The downside of the round-robin partitioning of instances is that it forces servers to commit requests at the same rate. For example, if server $s_1$ has not received a client request in time to start its Paxos instance 1, then $s_2$ would not be able to commit instance 2. Mencius addresses this problem by introducing coordinated paxoses for each instance in which each server proposes restricted values: only default leader can propose values for the corresponding instances, other servers can only propose no-op. If a default leader $s_j$ is suspected to have crashed, another server attempts to become the recovery leader by performing Phase-1 for instances in $I_j$. If the recovery leader finds out that the default leader has already proposed values for some instances, it will perform Phase-2 for those values. For all other instances, the recovery leader only does skips with no-ops. The nice property of coordinated-paxos is after the default leader proposes a no-op, we can learn no-op is going to be chosen without waiting for Phase-2. As a result, default leaders are able to skip their turns using no-op commands. In the earlier example, when $s_1$ receives the Phase-2 request from $s_2$ for instance 2, it piggybacks a skip message in its reply, thereby proposing no-op with instance 1. As soon as server $s_2$ receives the skip message, it can conclude that instance 1 is going to be committed with a no-op.

We can view Mencius as multiple multi-coordinated paxos groups, each group's instances share the same default leader. How to port the optimization of Mencius becomes to port multi-coordinated paxos to Raft. Figure 14 shows the algorithm based on MultiPaxos. First, each server maintains a boolean local variable isDefault which identifies if the server is the default leader or not. Second, each server has a boolean array skipsTags: skipsTags[i] is true if instance[i] has a no-op value proposed by the default leader. A server needs to attach this skipTags with its "prepareOk" message in Phase1b. After the leader collects the skipTags from $f + 1$ servers in Phase1Succeed , and uses these to update its local skipTags. When the leader propose a value, it also needs to attach "isDefault" to tell the follower if it is default leader or not. An acceptor will set the i-th entry in the skipTags if it accept "nop" value from the default leader and put it into an executableSet. All entries in the executableSet is executable even if it is not committed yet.

```
1  function Phase1b():
2    ...
3    reply <''prepareOK'', ..., skipTags>
4
5  function Phase1Succeed():
6    if receive <''prepareOK'', ..., skipTags> from f + 1 acceptors with the same b
7    ...
8        instance[i] = safeEntry(...)
9        m = message that has the safe value at i
10       skipsTags[i] = m.skipsTags[i]
11   ...
12
13 function Phase1c(i, v):
14   ...
15   send <''prePropose'', i, v, ballot, isDefault> to all
16
17 function Phase2a(i, v):
18   ...
19     send <''accept'', i, v, ballot, isDefault> to all
20
21 function Phase2b():
22   if receive <''accept'', i, v, b, default>
23     && b ≥ ballot
24   then
25     ...
26     if default && v == ''nop''
27     then
28       skipTags[i] = true
29       add <i, v> to executableSet
30     reply <''acceptOK'', i, v, b>
```

Fig. 14. Multi-coordinated paxos

### A.4 Raft[*]-Mencius

The complete version of Raft[*]-Mencius and other similar optimizations are included in the supplemental material. We only describe some interesting details here. In addition to the Paxos state variables, each replica needs to keep an array of "skip-tags", which indicates that which log entries can be skipped. When a replica becomes the leader, it needs to collect not only values but also skip-tags from other replicas. Because Phase2b action in Paxos corresponds to many actions ( AppendEntries, ReceiveAppend) in Raft[*], whatever changes Mencius makes to Phase2b should be applied to these actions as well. As an example, if the new appended entries are nop and from default leader, these entries should be marked executable.

Because Phase2b in Paxos is implied by multiple sub-actions in Raft[*], it is possible for a handworked solution in porting Mencius to miss some of the actions. For example, if the handworked solution only applies changes on Phase2b in Paxos to ReceiveAppend in Raft[*] (missing AppendEntries), the solution could miss some optimization opportunities or even generate an incorrect protocol.

```
1   function ReceiveVote():
2     ...
3     reply <''requestVoteOK'', ..., skipTags>
4
5   function BecomeLeader():
6     if receive <''requestVoteOK'', ..., skipTags > from f + 1 acceptors with the same t
7     ...
8        max = largest index of received entries
9        for i in lastIndex + 1 ... max
10           e = safeEntry(received entries with index i)
11           log[i].bal = currentTerm
12           log[i].term = currentTerm
13           log[i].val = e.val
14
15           m = message that has the safe value at i
16           skipsTags[i] = m.skipsTags[i]
17
18           if isDefault && e.val == ''nop''
19           then
20                skipTags[i] = true
21                add <i, e.val> to executableSet
22     ...
23
24  function AppendEntries(i, vals, prev):
25    ...
26       for each v in vals
27          n = lastIndex + 1
28          log[n].val = v
29          log[n].term = currentTerm
30          if isDefault && log[n].val == ''nop''
31          then
32               skipTags[n] = true
33               add <n, v> to executableSet
34     ...
35     send <''append'', currentTerm, prev, pTerm, ents,commitIndex, isDefault> to all
36
37  function ReceiveAppend():
38     if receive <''append'', t, prev, pTerm, ents, commit, default>
39        && t ≥ currentTerm
40     then
41        ...
42        for each entry e in ents
43           if default && e.val == ''nop''
44           then
45                skipTags[e.index] = true
46                add <e.index, e.val> to executableSet
47        ...
48        reply <''appendOK'', currentTerm, lastIndex>
```

Fig. 15. Coordinated Raft*

## B TLA[+] SPECIFICATIONS

We hereby provide the TLA[+] Specifications of the algorithms discussed in the paper, which includes : Paxos, Paxos*, Raft*, MBFT, PQL, Mencius and the ported optimizations on Raft* and MBFT.

### B.1 MultiPaxos

The TLA[+] Specification starts at the next page.

$\quad$ 1 $\vdash$ ————————— MODULE *MultiPaxos* —————————

Specification of *MultiPaxos*

5 EXTENDS *Integers*
6 $Min(s) \triangleq$ CHOOSE $x \in s : \forall\, y \in s : x \leq y$
7 $Max(s) \triangleq$ CHOOSE $x \in s : \forall\, y \in s : x \geq y$
8 $\vdash$ ————————————————————————————————————

10 CONSTANT *Quorum*,  The set of "quorums" where a quorum is a "large enough" set of acceptors
11 $\qquad\qquad$ *Value*,  The set of choosable values
12 $\qquad\qquad$ *Acceptor*  The of processes that will choose a value

14 VARIABLE *highestBallot*,  *highestBallot*[a] is the highest ballot number acceptor a has seen
15 $\qquad\qquad$ *isLeader*,  *isLeader*[a] is true if acceptor a is the leader, and can propose values
16 $\qquad\qquad$ *logTail*,  *logTail*[a] is just a pointer to show the end of the *log*
17 $\qquad\qquad$ *votes*,  *voes*[a][i] is the set of *votes* $< b, v >$ cast by acceptor a at index i
18 $\qquad\qquad$ *proposedValues*,  The set of proposed values at index i with ballot number b.
19 $\qquad\qquad$ *logs*,  *logs*[a][i] is latest vote $< b, v >$ *casted* by acceptor a at index i
20 $\qquad\qquad$ 1*amsgs*,
21 $\qquad\qquad$ 1*bmsgs*

23 $Ballot \triangleq Nat$
24 $Index \triangleq Nat$
25 $NoVal \triangleq$ CHOOSE $v : v \notin Value$
26 $\vdash$ ————————————————————————————————————

28 ASSUME $QuorumAssumption \triangleq \;\land\, \forall\, Q \in Quorum : Q \subseteq Quorum$
29 $\qquad\qquad\qquad\qquad\qquad\qquad\qquad\; \land\, \forall\, Q1,\, Q2 \in Quorum : Q1 \cap Q2 \neq \{\}$

31 $VotedFor(a,\, i,\, b,\, v) \triangleq \langle b,\, v \rangle \in votes[a][i]$

33 $ChosenAt(i,\, b,\, v) \triangleq \exists\, Q \in Quorum :$
34 $\qquad\qquad\qquad\qquad\quad \forall\, a \in Q : VotedFor(a,\, i,\, b,\, v)$

36 $chosen \triangleq [i \in Index \mapsto \{v \quad \in Value : \exists\, b \in Ballot : ChosenAt(i,\, b,\, v)\}]$

38 $DidNotVoteAt(a,\, i,\, b) \triangleq \forall\, v \in Value : \neg VotedFor(a,\, i,\, b,\, v)$

40 $CannotVoteAt(a,\, i,\, b) \triangleq\; \land\, highestBallot[a] > b$
41 $\qquad\qquad\qquad\qquad\qquad\quad\; \land\, DidNotVoteAt(a,\, i,\, b)$

43 $NoneOtherChoosableAt(i,\, b,\, v) \triangleq$
44 $\quad \exists\, Q \in Quorum :$
45 $\qquad \forall\, a \in Q : VotedFor(a,\, i,\, b,\, v) \lor CannotVoteAt(a,\, i,\, b)$

47 $SafeAt(i,\, b,\, v) \triangleq \forall\, c \in 0\,..\,(b-1) : NoneOtherChoosableAt(i,\, c,\, v)$

49 $ShowsSafeAt(Q,\, i,\, b,\, v) \triangleq$
50 $\quad \land\, \forall\, a \in Q : highestBallot[a] \geq b$
51 $\quad \land\, \exists\, c \in \quad -1\,..\,(b-1) :$

$$
\begin{aligned}
52 \quad & \wedge (c \neq -1) \Rightarrow \exists\, a \in Q : \mathit{VotedFor}(a,\, i,\, c,\, v) \\
53 \quad & \wedge \forall\, d \in (c+1)\,..\,(b-1),\, a \in Q : \mathit{DidNotVoteAt}(a,\, i,\, d)
\end{aligned}
$$

54 ⊢

56  $\mathit{OneValuePerBallot} \triangleq$
57    $\forall\, a1,\, a2 \in \mathit{Acceptor},\, b \in \mathit{Ballot},\, v1,\, v2 \in \mathit{Value},\, i \in \mathit{Index} :$
58    $\mathit{VotedFor}(a1,\, i,\, b,\, v1) \wedge \mathit{VotedFor}(a2,\, i,\, b,\, v2) \Rightarrow (v1 = v2)$

60  $\mathit{LogsSafe} \triangleq$
61    $\forall\, a \in \mathit{Acceptor},\, b \in \mathit{Ballot},\, i \in \mathit{Index},\, v \in \mathit{Value} :$
62    $(logs[a][i] = \langle b,\, v \rangle) \Rightarrow \mathit{SafeAt}(i,\, b,\, v)$

63 ⊢
64  $\mathit{GetHighestBallotEntry}(i,\, \mathit{logsIn1b}) \triangleq$
65    CHOOSE $bv \in ((\mathit{Ballot} \cup \{-1\}) \times (\mathit{Value} \cup \{\mathit{NoVal}\})) :$
66    $\wedge\quad \exists\, \mathit{log} \in \mathit{logsIn1b} : bv = \mathit{log}[i]$
67    $\wedge\quad \forall\, \mathit{log} \in \mathit{logsIn1b} : bv[1] \geq \mathit{log}[i][1]$

69  $\mathit{UpdateLog}(a,\, \mathit{logsIn1b},\, i1,\, i2) \triangleq$

> equivalent to :
> $\wedge\ logs' =$
>   $[\mathit{logs}\ \text{EXCEPT}\ ![a] =$
>     $[i \in \mathit{Index} \mapsto$
>       IF $\ i \in i1\,..\,i2\ $ THEN $\mathit{GetHighestBallotEntry}(i,\, \mathit{logsIn1b})$
>                               ELSE $logs[a][i]]]$
> $\wedge\ \mathit{logTail}' =$
>   IF $(i2 > \mathit{logTail}[a])$ THEN $[\mathit{logTail}\ \text{EXCEPT}\ ![a] = i2]$
>                                ELSE $\mathit{logTail}$

81    $\wedge\ \forall\, i \in i1\,..\,i2 : logs'[a][i] = \mathit{GetHighestBallotEntry}(i,\, \mathit{logsIn1b})$
82    $\wedge\ \forall\, x \in \mathit{Acceptor} \setminus \{a\} :$ UNCHANGED $logs[x]$
83    $\wedge\ \forall\, i \in \mathit{Index} \setminus i1\,..\,i2 :$ UNCHANGED $logs[a][i]$
84    $\wedge\ logs' \in [\mathit{Acceptor} \to [\mathit{Index} \to ((\mathit{Ballot} \cup \{-1\}) \times (\mathit{Value} \cup \{\mathit{NoVal}\}))]]$
85    $\wedge\ \mathit{logTail}' =$
86    $\quad$ IF $(i2 > \mathit{logTail}[a])$ THEN $[\mathit{logTail}\ \text{EXCEPT}\ ![a] = i2]$
87    $\qquad\qquad\qquad\qquad\ $ ELSE $\mathit{logTail}$

88 ⊢
89  $\mathit{IncreaseHighestBallot}(a,\, b) \triangleq$
90    $\wedge\ \mathit{highestBallot}[a] < b$
91    $\wedge\ \mathit{highestBallot}' = [\mathit{highestBallot}\ \text{EXCEPT}\ ![a] = b]$
92    $\wedge\ \mathit{isLeader}' = [\mathit{isLeader}\ \text{EXCEPT}\ ![a] = \text{FALSE}]$
93    $\wedge$ UNCHANGED $\langle \mathit{logTail},\, \mathit{votes},\, \mathit{proposedValues},\, \mathit{logs},\, 1\mathit{amsgs},\, 1\mathit{bmsgs} \rangle$

95  $\mathit{Phase1a}(a) \triangleq$
96    $\wedge\quad \neg \mathit{isLeader}[a]$
97    $\wedge\quad 1\mathit{amsgs}' = 1\mathit{amsgs} \cup \{[\mathit{acc} \mapsto a,\, \mathit{bal} \mapsto \mathit{highestBallot}[a]]\}$
98    $\wedge\quad$ UNCHANGED $\langle \mathit{highestBallot},\, \mathit{isLeader},\, \mathit{logTail},$
99    $\qquad\qquad\qquad\quad \mathit{votes},\, \mathit{proposedValues},\, \mathit{logs},\, 1\mathit{bmsgs} \rangle$

101    $Phase1b(a, 1amsg) \triangleq$
102      $\wedge$   $1amsg.bal > highestBallot[a]$
103      $\wedge$   $highestBallot' = [highestBallot \text{ EXCEPT } ![a] = 1amsg.bal]$
104      $\wedge$   $isLeader' = [isLeader \text{ EXCEPT } ![a] = \text{FALSE}]$
105      $\wedge$   $1bmsgs' = 1bmsgs \cup \{[acc \mapsto a, bal \mapsto 1amsg.bal,$
106         $log \mapsto logs[a], logTail \mapsto logTail[a]]\}$
107      $\wedge$   UNCHANGED $\langle logTail, votes, proposedValues, logs, 1amsgs \rangle$

109    $BecomeLeader(a, S) \triangleq$
110      $\wedge \neg isLeader[a]$
111      $\wedge \exists m \in S : m.acc = a$
112      $\wedge \forall m \in S : m.bal = highestBallot[a]$
113      $\wedge \{m.acc : m \in S\} \in Quorum$
114      $\wedge UpdateLog(a, \{m.log : m \in S\}, 0, Max(\{m.logTail : m \in S\}))$
115      $\wedge isLeader' = [isLeader \text{ EXCEPT } ![a] = \text{TRUE}]$
116      $\wedge$ UNCHANGED $\langle highestBallot, votes, proposedValues, 1amsgs, 1bmsgs \rangle$

118    $Propose(a, i, v) \triangleq$
119      $\wedge$   $isLeader[a]$
120      $\wedge$   $\vee \; logs[a][i][2] = v$
121         $\vee \; logs[a][i][2] = NoVal$   $\Rightarrow v$ is safe at $i, b$
122      $\wedge$   $proposedValues' = proposedValues \cup \{\langle i, highestBallot[a], v \rangle\}$
123      $\wedge$   UNCHANGED $\langle highestBallot, isLeader, logTail, votes, logs, 1amsgs, 1bmsgs \rangle$

125    $Accept(a, i, b, v) \triangleq$
126       $\wedge \langle i, b, v \rangle \in proposedValues$
127       $\wedge b \geq highestBallot[a]$
128       $\wedge highestBallot' = [highestBallot \text{ EXCEPT } ![a] = b]$
129       $\wedge votes' = [votes \text{ EXCEPT } ![a][i] = votes[a][i] \cup \{\langle b, v \rangle\}]$
130       $\wedge logs' = [logs \text{ EXCEPT } ![a][i] = \langle b, v \rangle]$
131       $\wedge logTail' = \text{IF } i > logTail[a] \text{ THEN } [logTail \text{ EXCEPT } ![a] = i] \text{ ELSE } logTail$
132       $\wedge isLeader' = \text{IF } b > highestBallot[a]$
133         THEN $[isLeader \text{ EXCEPT } ![a] = \text{FALSE}]$ ELSE $isLeader$
134       $\wedge$ UNCHANGED $\langle proposedValues, 1amsgs, 1bmsgs \rangle$

135 ⊢

136    $TypeOK \triangleq \;\; \wedge highestBallot \in [Acceptor \to Ballot]$
137           $\wedge isLeader \in [Acceptor \to \text{BOOLEAN}]$
138           $\wedge logTail \in [Acceptor \to Index \cup \{-1\}]$
139           $\wedge votes \in [Acceptor \to [Index \to \text{SUBSET }(Ballot \times Value)]]$
140           $\wedge proposedValues \in \text{SUBSET }(Index \times Ballot \times Value)$
141           $\wedge logs \in [Acceptor \to [Index \to (Ballot \cup \{-1\}) \times (Value \cup \{NoVal\})]]$
142           $\wedge 1amsgs \in \text{SUBSET }[acc : Acceptor, bal : Ballot]$
143           $\wedge 1bmsgs \in \text{SUBSET }[acc : Acceptor, bal : Ballot,$
144            $log : [Index \to (Ballot \cup \{-1\}) \times (Value \cup \{NoVal\})],$
145            $LogTail : Index \cup \{-1\}]$

$$
\begin{aligned}
&147 \quad Init \;\triangleq\; \wedge\, highestBallot = [a \in Acceptor \mapsto 0] \\
&148 \qquad\qquad \wedge\, isLeader = [a \in Acceptor \mapsto \text{FALSE}] \\
&149 \qquad\qquad \wedge\, logTail = [a \in Acceptor \mapsto -1] \\
&150 \qquad\qquad \wedge\, votes = [a \in Acceptor \mapsto [i \in Index \mapsto \{\}]] \\
&151 \qquad\qquad \wedge\, logs = [a \in Acceptor \mapsto [i \in Index \mapsto \langle -1,\, NoVal \rangle]] \\
&152 \qquad\qquad \wedge\, proposedValues = \{\} \\
&153 \qquad\qquad \wedge\, 1amsgs = \{\} \\
&154 \qquad\qquad \wedge\, 1bmsgs = \{\}
\end{aligned}
$$

$$
\begin{aligned}
&156 \quad Next \;\triangleq\; \vee\, \exists\, a \in Acceptor,\, b \in Ballot: \\
&157 \qquad\qquad\qquad IncreaseHighestBallot(a, b) \\
&158 \qquad\quad \vee\, \exists\, a \in Acceptor: \\
&159 \qquad\qquad\qquad Phase1a(a) \\
&160 \qquad\quad \vee\, \exists\, a \in Acceptor,\, m \in 1amsgs: \\
&161 \qquad\qquad\qquad Phase1b(a, m) \\
&162 \qquad\quad \vee\, \exists\, a \in Acceptor,\, S \in \text{SUBSET}\; 1bmsgs: \\
&163 \qquad\qquad\qquad BecomeLeader(a, S) \\
&164 \qquad\quad \vee\, \exists\, a \in Acceptor,\, i \in Index,\, v \in Value: \\
&165 \qquad\qquad\qquad Propose(a, i, v) \\
&166 \qquad\quad \vee\, \exists\, a \in Acceptor,\, pv \in proposedValues: \\
&167 \qquad\qquad\qquad Accept(a, pv[1], pv[2], pv[3])
\end{aligned}
$$

$$
\begin{aligned}
&169 \quad vars \;\triangleq\; \langle highestBallot,\, isLeader,\, logTail,\, votes, \\
&170 \qquad\qquad proposedValues,\, logs,\, 1amsgs,\, 1bmsgs \rangle
\end{aligned}
$$

$$
172 \quad Spec \;\triangleq\; Init \wedge \Box[Next]_{vars}
$$

$$
174 \quad Inv \;\triangleq\; TypeOK \wedge LogsSafe \wedge OneValuePerBallot
$$

176 ⊢──────────────────────────────────────────

177 THEOREM $Invariance \;\triangleq\; Spec \Rightarrow \Box Inv$

179 └──────────────────────────────────────────

### B.2 Raft*

The TLA$^+$ Specification starts at the next page.

1 ──────────────── MODULE $Raft*$ ────────────────

Specification of $Raft*$, whose leader will collect safe values in the empty $log$ entries when finishing phase 1.

6  EXTENDS $Integers$

8  CONSTANT $Quorum$,
9          $Value$,
10         $Acceptor$

12  VARIABLE $isLeader$,
13          $logTail$,
14          $lastIndex$,
15          $votes$,
16          $raftlogs$,
17          $proposedEntries$,
18          $highestBallot$,
19          $proposedValues$,
20          $logBallot$,
21          $r1amsgs$,
22          $r1bmsgs$

24  $Index \triangleq Nat$
25  $Ballot \triangleq Nat$

27  ASSUME $QuorumAssumption \triangleq \land \forall Q \in Quorum : Q \subseteq Quorum \land Q \subseteq Acceptor$
28                                     $\land \forall Q1, Q2 \in Quorum : Q1 \cap Q2 \neq \{\}$

31  $logs \triangleq [a \in Acceptor \mapsto$
32      $[i \in Index \mapsto \langle logBallot[a][i], raftlogs[a][i][2] \rangle]]$

34  $vars \triangleq \langle isLeader, logTail, lastIndex, raftlogs, logs,$
35  $votes, proposedValues, proposedEntries, highestBallot, logBallot, r1amsgs, r1bmsgs \rangle$
36  ├─────────────────────────────────────────────────────────────────────────
37  $1amsgs \triangleq \{[acc \mapsto m.acc, bal \mapsto m.bal] : m \in r1amsgs\}$
38  $1bmsgs \triangleq r1bmsgs$

40  $MP \triangleq$ INSTANCE $MultiPaxos$
41  $Max(x) \triangleq MP!Max(x)$
42  $NoVal \triangleq MP!NoVal$
43  ├─────────────────────────────────────────────────────────────────────────
44  $TypeOK \triangleq \land highestBallot \in [Acceptor \to Ballot]$
45               $\land isLeader \in [Acceptor \to \text{BOOLEAN}]$
46               $\land lastIndex \in [Acceptor \to Index \cup \{-1\}]$
47               $\land logTail \in [Acceptor \to Index \cup \{-1\}]$
48               $\land votes \in [Acceptor \to [Index \to \text{SUBSET}(Ballot \times Value)]]$
49               $\land raftlogs \in [Acceptor \to [Index \to$

```
50              ((Ballot ∪ { − 1}) × (Value ∪ {NoVal}))]]
51          ∧ logs ∈ [Acceptor → [Index →
52              ((Ballot ∪ { − 1}) × (Value ∪ {NoVal}))]]
53          ∧ proposedEntries ∈ SUBSET [term : Ballot,
54                                      prevLogTerm : Ballot ∪ { − 1},
55                                      prevLogIndex : Index ∪ { − 1},
56                                      lIndex : Index ∪ { − 1},
57                                      leaderId : Acceptor,
58                                      leaderCommit : Index ∪ { − 1},
59                                      entries : [Index → (Ballot × Value)]
60                                     ]
61          ∧ logBallot ∈ [Acceptor → [Index → Ballot ∪ { − 1}]]
62          ∧ proposedValues ∈ SUBSET (Index × Ballot × Value)
63          ∧ r1amsgs ∈ SUBSET [acc : Acceptor, bal : Ballot,
64              lastTerm : Ballot ∪ { − 1}, lastIndex : Index ∪ { − 1}]
65          ∧ r1bmsgs ∈ SUBSET [acc : Acceptor, bal : Ballot,
66              log : [Index → (Ballot ∪ { − 1}) × (Value ∪ {NoVal})]]

68 ⊢─────────────────────────────────────────────────────────────────────────
69 UpdateLog(a, logsIn1B, i1, i2) ≜
70      This can Imply : logs′[a][i] = MP!GetHighestBallotEntry(i, logsIn1B)
71      ∧ ∀ i ∈ i1 .. i2 :
72          ∧ raftlogs′[a][i] = ⟨ − 1, MP!GetHighestBallotEntry(i, logsIn1B)[2]⟩
73          ∧ logBallot′[i] = MP!GetHighestBallotEntry(i, logsIn1B)[1]
74      This can Imply : i ∈ Index \ i1 .. i2 : UNCHANGED logs[a][i]
75      ∧ ∀ i ∈ Index \ i1 .. i2 : UNCHANGED ⟨raftlogs[a][i], logBallot[a][i]⟩
76      This can Imply : x ∈ Acceptor \ {a} : UNCHANGED logs[x]
77      ∧ ∀ x ∈ Acceptor \ {a} : UNCHANGED ⟨raftlogs[x], logBallot[x]⟩
78      ∧ raftlogs′ ∈ [Acceptor →
79          [Index → ((Ballot ∪ { − 1}) × (Value ∪ {NoVal}))]]
80      ∧ logTail′ =
81          IF (i2 > logTail[a]) THEN [logTail EXCEPT ![a] = i2] ELSE logTail
82 ⊢─────────────────────────────────────────────────────────────────────────
83 IncreaseHighestBallot(a, b) ≜
84      ∧ MP!IncreaseHighestBallot(a, b)
85      ∧ UNCHANGED ⟨isLeader, lastIndex, logTail, logs, raftlogs,
86          logBallot, votes, proposedValues, proposedEntries⟩

88 Phase1a(a) ≜
89      ∧   r1amsgs′ = r1amsgs ∪ {[acc ↦ a,
90                                 bal ↦ highestBallot[a],
91                                 lastTerm ↦ IF lastIndex[a] = − 1 THEN − 1
92                                            ELSE raftlogs[a][lastIndex[a]][1],
93                                 lastIndex ↦ lastIndex[a]]}
94      ∧   UNCHANGED ⟨highestBallot, isLeader, lastIndex, logTail, logs,
```

```
 95              raftlogs, logBallot, votes, proposedValues, proposedEntries, r1bmsgs⟩

 97   Phase1b(a, r1amsg) ≜
 98       ∧  r1amsg.bal > highestBallot[a]
 99       ∧  ∨ lastIndex[a] = −1
100          ∨ ∧ lastIndex[a] ≠ −1
101            ∧ raftlogs[a][lastIndex[a]][1] < r1amsg.lastTerm
102          ∨ ∧ lastIndex[a] ≠ −1
103            ∧ raftlogs[a][lastIndex[a]][1] = r1amsg.lastTerm
104            ∧ lastIndex[a] ≤ r1amsg.lastIndex
105       ∧  highestBallot' = [highestBallot EXCEPT ![a] = r1amsg.bal]
106       ∧  isLeader' = [isLeader EXCEPT ![a] = FALSE]
107       ∧  r1bmsgs' = r1bmsgs ∪ {[acc ↦ a, bal ↦ r1amsg.bal,
108          log ↦ logs[a], logTail ↦ logTail[a]]}
109       ∧  UNCHANGED ⟨lastIndex, logTail, logs, raftlogs, logBallot,
110          votes, proposedValues, proposedEntries, r1amsgs⟩

112   BecomeLeader(a, S) ≜
113       ∧ ∃ m ∈ S : m.acc = a
114       ∧ ∀ m ∈ S : m.bal = highestBallot[a]
115       ∧ {m.acc : m ∈ S} ∈ Quorum
116       ∧ ∀ i ∈ 0 .. lastIndex[a] : UNCHANGED ⟨logBallot[a][i], raftlogs[a][i]⟩
117       the above 4 conditions ⇒ UpdateLog(a, {m.log : m ∈ S}, 0, lastIndex[a])
118       ∧ UpdateLog(a, {m.log : m ∈ S}, lastIndex[a] + 1, Max({m.logTail : m ∈ S}))
119       ∧ isLeader' = [isLeader EXCEPT ![a] = TRUE]
120       ∧ UNCHANGED ⟨highestBallot, lastIndex, votes, proposedValues,
121          proposedEntries, r1bmsgs, r1amsgs⟩

123   ProposeEntries(a, i1, i, v) ≜
124       ∧ isLeader[a]
125       ∧ i = logTail[a] + 1
126       ∧ proposedEntries' = proposedEntries ∪
127          [term ↦ highestBallot[a], prevLogTerm ↦
128              IF i1 = 0 THEN raftlogs[a][i1 − 1][1] ELSE  −1,
129          prevLogIndex ↦ i1 − 1, lIndex ↦ i, leaderId ↦ a,
130          leaderCommit ↦ −1, entries ↦ [j ∈ i1 .. i ↦
131              IF j = i THEN ⟨highestBallot[a], v⟩ ELSE  raftlogs[a][j]]]
132       ∧ proposedValues' = proposedValues ∪ {ibv ∈ (Index × Ballot × Value) :
133          raftlogs[a][ibv[1]][2] = ibv[3] ∧ ibv[2]      = highestBallot[a] ∧ ibv[1] ∈ 0 .. i}
134       ∧ UNCHANGED ⟨isLeader, lastIndex, logTail, raftlogs, logs, logBallot,
135          votes, highestBallot, r1amsgs, r1bmsgs⟩

137   AcceptEntries(a, pe) ≜
138       ∧ pe.term ≥ highestBallot[a]
```

> We have two methods to preserve *LogMatchingInv*. The first one is to force the acceptors only accept longer logs, thus it can replace all its old entries. The second method is to delete the entries with old terms. However, paxos doesn't provide delete operation, so we can not map this state to paxos. So we choose the first method.

$$
\begin{aligned}
&146 \quad \land\ pe.lIndex \geq lastIndex[a] \\
&147 \quad \land\ (pe.prevLogIndex > -1) \Rightarrow (raftlogs[a][pe.prevLogIndex][1] = pe.prevLogTerm) \\
&148 \quad \land\ \forall\, i \in 0\,..\,pe.lIndex : logBallot'[a][i] = pe.term \\
&149 \quad \land\ \forall\, i \in pe.prevLogIndex + 1\,..\,pe.lIndex : \\
&150 \qquad \land\ raftlogs'[a][i] = pe.entries[i] \\
&151 \qquad \land\ votes'[a][i] = votes[a][i]\ \cup \\
&152 \qquad\quad \{\langle pe.entries[pe.lIndex][1],\ pe.entries[i][2]\rangle\} \\
&153 \quad \land\ \forall\, i \in Index \setminus (0\,..\,pe.lIndex) : \\
&154 \qquad \text{UNCHANGED}\ \langle raftlogs[a][i],\ votes[a][i],\ logBallot[a][i]\rangle \\
&155 \quad \land\ \forall\, x \in Acceptor \setminus \{a\} : \\
&156 \qquad \text{UNCHANGED}\ \langle raftlogs[x],\ votes[x],\ logBallot[x]\rangle \\
&157 \quad \land\ highestBallot' = [highestBallot\ \text{EXCEPT}\ ![a] = pe.term] \\
&158 \quad \land\ lastIndex' = \text{IF}\ pe.lIndex > lastIndex[a] \\
&159 \qquad \text{THEN}\ [lastIndex\ \text{EXCEPT}\ ![a] = pe.lIndex]\ \text{ELSE}\ lastIndex \\
&160 \quad \land\ logTail' = \text{IF}\ pe.lIndex > logTail[a] \\
&161 \qquad \text{THEN}\ [logTail\ \text{EXCEPT}\ ![a] = pe.lIndex]\ \text{ELSE}\ logTail \\
&162 \quad \land\ isLeader' = \text{IF}\ pe.term > highestBallot[a] \\
&163 \qquad \text{THEN}\ [isLeader\ \text{EXCEPT}\ ![a] = \text{FALSE}]\ \text{ELSE}\ isLeader \\
&164 \quad \land\ (pe.term > highestBallot[a]) \Rightarrow (isLeader' = [isLeader\ \text{EXCEPT}\ ![a] = \text{FALSE}]) \\
&165 \quad \land\ \text{UNCHANGED}\ \langle proposedEntries,\ proposedValues,\ r1amsgs,\ r1bmsgs\rangle
\end{aligned}
$$

$$
\begin{aligned}
&167\ Init\ \triangleq\ \land\ highestBallot = [a \in Acceptor \mapsto 0] \\
&168 \qquad\qquad \land\ isLeader = [a \in Acceptor \mapsto \text{FALSE}] \\
&169 \qquad\qquad \land\ logTail = [a \in Acceptor \mapsto -1] \\
&170 \qquad\qquad \land\ lastIndex = [a \in Acceptor \mapsto -1] \\
&171 \qquad\qquad \land\ votes = [a \in Acceptor \mapsto [i \in Index \mapsto \{\}]] \\
&172 \qquad\qquad \land\ logs = [a \in Acceptor \mapsto [i \in Index \mapsto \langle -1,\ NoVal\rangle]] \\
&173 \qquad\qquad \land\ raftlogs = [a \in Acceptor \mapsto [i \in Index \mapsto \langle -1,\ NoVal\rangle]] \\
&174 \qquad\qquad \land\ logBallot = [a \in Acceptor \mapsto [i \in Index \mapsto -1]] \\
&175 \qquad\qquad \land\ proposedEntries = \{\} \\
&176 \qquad\qquad \land\ proposedValues = \{\} \\
&177 \qquad\qquad \land\ r1amsgs = \{\} \\
&178 \qquad\qquad \land\ r1bmsgs = \{\}
\end{aligned}
$$

$$
\begin{aligned}
&180\ Next\ \triangleq\ \lor\ \exists\, a \in Acceptor,\ b \in Ballot : \\
&181 \qquad\qquad IncreaseHighestBallot(a, b) \\
&182 \qquad \lor\ \exists\, a \in Acceptor : \\
&183 \qquad\qquad Phase1a(a) \\
&184 \qquad \lor\ \exists\, a \in Acceptor,\ m \in r1amsgs : \\
&185 \qquad\qquad Phase1b(a, m) \\
&186 \qquad \lor\ \exists\, a \in Acceptor,\ S \in \text{SUBSET}\ r1bmsgs : \\
&187 \qquad\qquad BecomeLeader(a, S)
\end{aligned}
$$

```
188             ∨ ∃ a, x ∈ Acceptor, i1, i ∈ Index, v ∈ Value :
189                 ProposeEntries(a, i1, i, v)
190             ∨ ∃ a ∈ Acceptor, pe ∈ proposedEntries :
191                 AcceptEntries(a, pe)

193   Spec  ≜  Init ∧ □[Next]_vars
```

---

```
196   LogMatchingInv  ≜
197       ∀ x ∈ Acceptor, y ∈ Acceptor, i ∈ Index :
198           raftlogs[x][i][1] = raftlogs[y][i][1] ⇒
199               ∀ j ∈ 0 .. i : raftlogs[x][j] = raftlogs[y][j]

201   LeaderCompletenessInv  ≜
202       ∀ i ∈ Index, b ∈ Ballot, v ∈ Value :
203           (∃ Q ∈ Quorum : ∀ x ∈ Q : raftlogs[x][i] = ⟨b, v⟩)
204           ⇒
205           (∀ a ∈ Acceptor, S ∈ SUBSET r1bmsgs :
206               (ENABLED BecomeLeader(a, S)) ⇒ raftlogs[a][i] = ⟨b, v⟩)

208   LogBallotInv  ≜
209       ∀ a ∈ Acceptor :
210           ∧ ∀ i ∈ 0 .. lastIndex[a] :
211             logBallot[a][i] = raftlogs[a][lastIndex[a]][1]
212           ∧ ∀ i ∈ Index \ 0 .. lastIndex[a] :
213             (lastIndex[a] ≠ −1) ⇒
214             (logBallot[a][i] ≤ raftlogs[a][lastIndex[a]][1])
215           ∧ lastIndex[a] = −1 ⇒ logTail[a] = −1

218   RaftInv  ≜  ∧ TypeOK
219                ∧ LogMatchingInv   this property was proved in Raft paper
220                ∧ LeaderCompletenessInv   safety property, also proved in Raft paper
221                ∧ LogBallotInv

223   Inv  ≜  MP!Inv ∧ RaftInv

225   THEOREM  Invariance  ≜  Spec ⇒ □Inv
```

## B.3 Paxos Quorum Lease

The TLA$^+$ Specification starts at the next page.

```
 1 ┌──────────────── MODULE PQL ────────────────
   Specification of Paxos Quorum Lease. For the unmodified subactions from MultiPaxos, this spec
   directly use the subactions of MP.
 6 EXTENDS Integers
 7 Min(s) ≜ CHOOSE x ∈ s : ∀ y ∈ s : x ≤ y
 8 Max(s) ≜ CHOOSE x ∈ s : ∀ y ∈ s : x ≥ y
 9 ├─────────────────────────────────────────────

11 CONSTANT Quorum,
12          Value,
13          Acceptor

15 VARIABLE highestBallot, isLeader, logTail, votes,
16          proposedValues, logs, applyIndex, 1amsgs, 1bmsgs

18 Ballot ≜ Nat
19 Index  ≜ Nat
20 NoVal  ≜ CHOOSE v : v ∉ Value
21 ├─────────────────────────────────────────────

23   Constant and Variables for PQL
24 CONSTANT LeaseDuration

26   the distributed lease protocal implements this simple lease protocol with a global timer
27 VARIABLE timer,    assume there is a global timer
28          leases    grantedLeases[p][q] is the lease information < deadline > granted by p to q

30 PQLvars ≜ ⟨applyIndex, timer, leases⟩
31 ├─────────────────────────────────────────────
32 ASSUME QuorumAssumption ≜ ∧ ∀ Q ∈ Quorum : Q ⊆ Quorum
33                           ∧ ∀ Q1, Q2 ∈ Quorum : Q1 ∩ Q2 ≠ {}

35 ASSUME ValueTypeAssumption ≜ ∀ v ∈ Value :
36    ∨  v = NoVal
37    ∨  v.type = "read"
38    ∨  v.type = "write"

40 VotedFor(a, i, b, v) ≜ ⟨b, v⟩ ∈ votes[a][i]

42 ChosenAt(i, b, v) ≜ ∃ Q ∈ Quorum :
43                        ∀ a ∈ Q : VotedFor(a, i, b, v)

45 chosen ≜ [i ∈ Index ↦ {v ∈ Value : ∃ b ∈ Ballot : ChosenAt(i, b, v)}]

47 ├─────────────────────────────────────────────
48   Here we derive PQL by modifying MultiPaxos
49 MP ≜ INSTANCE MultiPaxos

51   The lease related state functions and subactions
```

52    $LeaseIsActive(p) \triangleq \exists\, Q \in Quorum : \forall\, a \in Q : leases[a][p] \geq timer$

54    $ActiveLeaseHolders \triangleq \{p \in Acceptor : LeaseIsActive(p)\}$

56    $GrantedLeaseHolders(Q) \triangleq \{p \in Acceptor : \exists\, a \in Q : leases[a][p] \geq timer\}$

58    a value can be executable only if all the lease holders have acknowledged this change
59    $CanCommitAt(i,\, b,\, v) \triangleq$
60       $\exists\, Q \in Quorum :$
61          $\land\, \forall\, a \in Q : VotedFor(a,\, i,\, b,\, v)$    $v$ is chosen by $Q$
62          $\land\, \forall\, p \in GrantedLeaseHolders(Q) : VotedFor(p,\, i,\, b,\, v)$
63        $v$ is known by every lease holder granted by $\forall\, a \in Q$
64        $\Rightarrow v$ is known by every active lease holder (Proved by the property of $Quorum$)

66    $executable \triangleq \{ibv \in Index \times Ballot \times Value : CanCommitAt(ibv[1],\, ibv[2],\, ibv[3])\}$

Here we add the apply subaction to do optimizations related to real implementations.

73    $Apply(a,\, i) \triangleq$
74       $\land\, i = applyIndex[a] + 1$
75       $\land\, CanCommitAt(i,\, logs[a][i][1],\, logs[a][i][2])$
76       $\land\, applyIndex' = [applyIndex \text{ EXCEPT } ![a] = i]$
77       $\land\, \text{UNCHANGED } MP!vars$

79    $p$ grant lease to $q$
80    $GrantLease(p,\, q) \triangleq$
81       $\land\, leases' = [leases \text{ EXCEPT } ![p][q] = timer + LeaseDuration]$
82       $\land\, \text{UNCHANGED } MP!vars$

84    $UpdateTimer \triangleq$
85       $\land\, timer' = timer + 1$
86       $\land\, \text{UNCHANGED } MP!vars$

a value is readable only if all the entries which update the value are in the executable set and have been applied

92    $ReadAtLocal(a) \triangleq$
93       $\land\, LeaseIsActive(a)$
94       $\land\, logTail[a] = applyIndex[a]$    wait for all pending write commands to finish
95       $\land\, \text{TRUE}$   Do local reading (does not change any server state)
96       $\land\, \text{UNCHANGED } \langle MP!vars,\, PQLvars \rangle$

98    $executableEntryAt(i) \triangleq$
99       IF $\exists\, ibv \in executable : ibv[1] = i$
100          THEN CHOOSE $ibv \in executable : ibv[1] = i$
101          ELSE "No"

103    $TypeOK \triangleq\ \land\, timer \in Nat$
104               $\land\, leases \in [Acceptor \rightarrow [Acceptor \rightarrow Nat]]$

```
105            ∧ LeaseDuration ∈ Nat
106            ∧ applyIndex ∈ [Acceptor → Index ∪ {−1}]

108  Next ≜  ∨ ∃ a ∈ Acceptor, b ∈ Ballot :
109              MP!IncreaseHighestBallot(a, b) ∧ UNCHANGED PQLvars
110           ∨ ∃ a ∈ Acceptor :
111              MP!Phase1a(a) ∧ UNCHANGED PQLvars
112           ∨ ∃ a ∈ Acceptor, Q ∈ Quorum :
113              MP!BecomeLeader(a, Q) ∧ UNCHANGED PQLvars
114           ∨ ∃ a ∈ Acceptor, i ∈ Index, v ∈ Value :
115              ∧ ∨ v.type ≠ "read"
116                 ∨ ¬LeaseIsActive(a)
117              ∧ MP!Propose(a, i, v)
118              ∧ UNCHANGED PQLvars
119           ∨ ∃ a ∈ Acceptor, pv ∈ proposedValues :
120              MP!Accept(a, pv[1], pv[2], pv[3]) ∧ UNCHANGED PQLvars
121           ∨ ∃ a ∈ Acceptor, i ∈ Index :
122              Apply(a, i)
123           ∨ ∃ a, p ∈ Acceptor :
124              GrantLease(a, p)
125           ∨ ∃ a ∈ Acceptor :
126              ReadAtLocal(a)
127           ∨ UpdateTimer
128  ├──────────────────────────────────────────────────────────────

130  LeaseSafe(i, b, v) ≜  ∧ ChosenAt(i, b, v)  A executable value must have been chosen
131                        ∧ ∀ a ∈ ActiveLeaseHolders : VotedFor(a, i, b, v)
132                           All lease holders knows this value.

134  linearizability: all read / write actions are totally ordered
135  LeaseInv ≜ ∀ ⟨i, b, v⟩ ∈ executable : LeaseSafe(i, b, v)
136  └──────────────────────────────────────────────────────────────
```

### B.4 Raft*-PQL

The TLA$^+$ Specification starts at the next page.

```
 1 ┌─────────────────── MODULE RQL ───────────────────
   Specification of Raft*-PQL
 5 EXTENDS Integers

 7 CONSTANT Quorum,
 8          Value,
 9          Acceptor

11 VARIABLE isLeader,
12          logTail,
13          lastIndex,
14          votes,
15          raftlogs,
16          proposedEntries,
17          highestBallot,
18          proposedValues,
19          logBallot,
20          r1amsgs,
21          r1bmsgs,
22          applyIndex

24 Index ≜ Nat
25 Ballot ≜ Nat

27 ASSUME QuorumAssumption ≜ ∧ ∀ Q ∈ Quorum : Q ⊆ Quorum ∧ Q ⊆ Acceptor
28                           ∧ ∀ Q1, Q2 ∈ Quorum : Q1 ∩ Q2 ≠ {}

31 logs ≜ [a ∈ Acceptor ↦ [i ∈ Index ↦ ⟨logBallot[a][i], raftlogs[a][i][2]⟩]]

33 vars ≜ ⟨isLeader, logTail, lastIndex, raftlogs, logs, votes,
34        proposedValues, proposedEntries, highestBallot, logBallot, r1amsgs, r1bmsgs⟩
35 ├─────────────────────────────────────────────────
36 1amsgs ≜ {[acc ↦ m.acc, bal ↦ m.bal] : m ∈ r1amsgs}
37 1bmsgs ≜ r1bmsgs

39 MP ≜ INSTANCE MultiPaxosM

41 Max(x) ≜ MP!Max(x)
42 NoVal ≜ MP!NoVal
43 ├─────────────────────────────────────────────────
44 CONSTANT LeaseDuration

46     the distributed lease protocal implements this simple lease protocal with a global timer
47 VARIABLE timer,    assume there is a global timer
48          leases    grantedLeases[p][q] is the lease information < deadline > granted by p to q

50 RQLvars ≜ ⟨applyIndex, timer, leases⟩
```

$\vdash$

$TypeOKRQL \triangleq \land timer \in Nat$
$\phantom{TypeOKRQL \triangleq} \land leases \in [Acceptor \to [Acceptor \to Nat]]$
$\phantom{TypeOKRQL \triangleq} \land LeaseDuration \in Nat$
$\phantom{TypeOKRQL \triangleq} \land applyIndex \in [Acceptor \to Index \cup \{-1\}]$

$TypeOK \triangleq \land highestBallot \in [Acceptor \to Ballot]$
$\phantom{TypeOK \triangleq} \land isLeader \in [Acceptor \to \text{BOOLEAN}]$
$\phantom{TypeOK \triangleq} \land lastIndex \in [Acceptor \to Index \cup \{-1\}]$
$\phantom{TypeOK \triangleq} \land logTail \in [Acceptor \to Index \cup \{-1\}]$
$\phantom{TypeOK \triangleq} \land votes \in [Acceptor \to [Index \to \text{SUBSET}\,(Ballot \times Value)]]$
$\phantom{TypeOK \triangleq} \land raftlogs \in [Acceptor \to$
$\phantom{TypeOK \triangleq \land} [Index \to ((Ballot \cup \{-1\}) \times (Value \cup \{NoVal\}))]]$
$\phantom{TypeOK \triangleq} \land logs \in [Acceptor \to$
$\phantom{TypeOK \triangleq \land} [Index \to ((Ballot \cup \{-1\}) \times (Value \cup \{NoVal\}))]]$
$\phantom{TypeOK \triangleq} \land proposedEntries \in \text{SUBSET}\,[term : Ballot,$
$\phantom{TypeOK \triangleq \land proposedEntries \in} prevLogTerm : Ballot \cup \{-1\},$
$\phantom{TypeOK \triangleq \land proposedEntries \in} prevLogIndex : Index \cup \{-1\},$
$\phantom{TypeOK \triangleq \land proposedEntries \in} lIndex : Index \cup \{-1\},$
$\phantom{TypeOK \triangleq \land proposedEntries \in} leaderId : Acceptor,$
$\phantom{TypeOK \triangleq \land proposedEntries \in} leaderCommit : Index \cup \{-1\},$
$\phantom{TypeOK \triangleq \land proposedEntries \in} entries : [Index \to (Ballot \times Value)]$
$\phantom{TypeOK \triangleq \land proposedEntries \in} ]$
$\phantom{TypeOK \triangleq} \land logBallot \in [Acceptor \to [Index \to Ballot \cup \{-1\}]]$
$\phantom{TypeOK \triangleq} \land proposedValues \in \text{SUBSET}\,(Index \times Ballot \times Value)$
$\phantom{TypeOK \triangleq} \land r1amsgs \in \text{SUBSET}\,[acc : Acceptor, bal : Ballot,$
$\phantom{TypeOK \triangleq \land} lastTerm : Ballot \cup \{-1\}, lastIndex : Index \cup \{-1\}]$
$\phantom{TypeOK \triangleq} \land r1bmsgs \in \text{SUBSET}\,[acc : Acceptor, bal : Ballot,$
$\phantom{TypeOK \triangleq \land} log : [Index \to (Ballot \cup \{-1\}) \times (Value \cup \{NoVal\})]]$
$\phantom{TypeOK \triangleq} \land TypeOKRQL$

$\vdash$

$UpdateLog(a, logsIn1B, i1, i2) \triangleq$
$\phantom{UpdateLog}$ This can Imply : $logs'[a][i] = MP!GetHighestBallotEntry(i, logsIn1B)$
$\phantom{UpdateLog} \land \forall i \in i1\,..\,i2 :$
$\phantom{UpdateLog \land \forall} \land raftlogs'[a][i] = \langle -1, MP!GetHighestBallotEntry(i, logsIn1B)[2] \rangle$
$\phantom{UpdateLog \land \forall} \land logBallot'[i] = MP!GetHighestBallotEntry(i, logsIn1B)[1]$
$\phantom{UpdateLog}$ This can Imply : $i \in Index \setminus i1\,..\,i2 : \text{UNCHANGED}\ logs[a][i]$
$\phantom{UpdateLog} \land \forall i \in Index \setminus i1\,..\,i2 : \text{UNCHANGED}\ \langle raftlogs[a][i], logBallot[a][i] \rangle$
$\phantom{UpdateLog}$ This can Imply : $x \in Acceptor \setminus \{a\} : \text{UNCHANGED}\ logs[x]$
$\phantom{UpdateLog} \land \forall x \in Acceptor \setminus \{a\} : \text{UNCHANGED}\ \langle raftlogs[x], logBallot[x] \rangle$
$\phantom{UpdateLog} \land raftlogs' \in [Acceptor \to [Index \to ((Ballot \cup \{-1\}) \times (Value \cup \{NoVal\}))]]$
$\phantom{UpdateLog} \land logTail' =$
$\phantom{UpdateLog \land} \text{IF}\ (i2 > logTail[a])\ \text{THEN}\ [logTail\ \text{EXCEPT}\ ![a] = i2]\ \text{ELSE}\ logTail$

$\vdash$

$IncreaseHighestBallot(a, b) \triangleq$

```
 97        ∧ MP!IncreaseHighestBallot(a, b)
 98        ∧ UNCHANGED ⟨isLeader, lastIndex, logTail, logs, raftlogs,
 99            logBallot, votes, proposedValues, proposedEntries⟩

101   Phase1a(a) ≜
102       ∧  r1amsgs' = r1amsgs ∪ {[acc ↦ a,
103                                 bal ↦ highestBallot[a],
104                                 lastTerm ↦ IF lastIndex[a] = −1 THEN −1
105                                              ELSE raftlogs[a][lastIndex[a]][1],
106                                 lastIndex ↦ lastIndex[a]]}
107       ∧  UNCHANGED ⟨highestBallot, isLeader, lastIndex, logTail, logs,
108            raftlogs, logBallot, votes, proposedValues, proposedEntries, r1bmsgs⟩

110   Phase1b(a, r1amsg) ≜
111       ∧  r1amsg.bal > highestBallot[a]
112       ∧  ∨ lastIndex[a] = −1
113          ∨ ∧ lastIndex[a] ≠ −1
114            ∧ raftlogs[a][lastIndex[a]][1] < r1amsg.lastTerm
115          ∨ ∧ lastIndex[a] ≠ −1
116            ∧ raftlogs[a][lastIndex[a]][1] = r1amsg.lastTerm
117            ∧ lastIndex[a] ≤ r1amsg.lastIndex
118       ∧  highestBallot' = [highestBallot EXCEPT ![a] = r1amsg.bal]
119       ∧  isLeader' = [isLeader EXCEPT ![a] = FALSE]
120       ∧  r1bmsgs' = r1bmsgs ∪ {[acc ↦ a, bal ↦ r1amsg.bal,
121            log ↦ logs[a], logTail ↦ logTail[a]]}
122       ∧  UNCHANGED ⟨lastIndex, logTail, logs, raftlogs, logBallot,
123            votes, proposedValues, proposedEntries, r1amsgs⟩

125   BecomeLeader(a, S) ≜
126       ∧ ∃ m ∈ S : m.acc = a
127       ∧ ∀ m ∈ S : m.bal = highestBallot[a]
128       ∧ {m.acc : m ∈ S} ∈ Quorum
129       ∧ ∀ i ∈ 0 .. lastIndex[a] : UNCHANGED ⟨logBallot[a][i], raftlogs[a][i]⟩
130       the above 4 conditions ⇒ UpdateLog(a, {m.log : m ∈ S}, 0, lastIndex[a])
131       ∧ UpdateLog(a, {m.log : m ∈ S}, lastIndex[a] + 1, Max({m.logTail : m ∈ S}))
132       ∧ isLeader' = [isLeader EXCEPT ![a] = TRUE]
133       ∧ UNCHANGED ⟨highestBallot, lastIndex, votes, proposedValues,
134            proposedEntries, r1bmsgs, r1amsgs⟩

136   ProposeEntries(a, i1, i, v) ≜
137       ∧ isLeader[a]
138       ∧ i = logTail[a] + 1
139       ∧ proposedEntries' = proposedEntries ∪
140         [term ↦ highestBallot[a], prevLogTerm ↦
141            IF i1 = 0 THEN raftlogs[a][i1 − 1][1] ELSE  − 1,
142         prevLogIndex ↦ i1 − 1, lIndex ↦ i, leaderId ↦ a,
```

```
143          leaderCommit ↦ −1, entries ↦ [j ∈ i1 . . i  ↦
144              IF j = i THEN ⟨highestBallot[a], v⟩ ELSE raftlogs[a][j]]]
145      ∧ proposedValues′ = proposedValues ∪ {ibv ∈ (Index × Ballot × Value) :
146          raftlogs[a][ibv[1]][2] = ibv[3] ∧ ibv[2]    = highestBallot[a] ∧ ibv[1] ∈ 0 . . i}
147      ∧ UNCHANGED ⟨isLeader, lastIndex, logTail, raftlogs, logs, logBallot,
148          votes, highestBallot, r1amsgs, r1bmsgs⟩

150  AcceptEntries(a, pe) ≜
151      ∧ pe.term ≥ highestBallot[a]
```

> We have two methods to preserve *LogMatchingInv*. The first one is to force the acceptors only accept longer logs, thus it can replace all its old entries. The second method is to delete the entries with old terms. However, paxos doesn't provide delete operation, so we can not map this state to paxos. So we choose the first method.

```
159      ∧ pe.lIndex ≥ lastIndex[a]
160      ∧ (pe.prevLogIndex > −1) ⇒ (raftlogs[a][pe.prevLogIndex][1] = pe.prevLogTerm)
161      ∧ ∀ i ∈ 0 . . pe.lIndex : logBallot′[a][i] = pe.term
162      ∧ ∀ i ∈ pe.prevLogIndex + 1 . . pe.lIndex :
163          ∧ raftlogs′[a][i] = pe.entries[i]
164          ∧ votes′[a][i] = votes[a][i] ∪ {⟨pe.entries[pe.lIndex][1], pe.entries[i][2]⟩}
165      ∧ ∀ i ∈ Index \ (0 . . pe.lIndex) :
166          UNCHANGED ⟨raftlogs[a][i], votes[a][i], logBallot[a][i]⟩
167      ∧ ∀ x ∈ Acceptor \ {a} :
168          UNCHANGED ⟨raftlogs[x], votes[x], logBallot[x]⟩
169      ∧ highestBallot′ = [highestBallot EXCEPT ![a] = pe.term]
170      ∧ lastIndex′ = IF pe.lIndex > lastIndex[a]
171          THEN [lastIndex EXCEPT ![a] = pe.lIndex] ELSE lastIndex
172      ∧ logTail′ = IF pe.lIndex > logTail[a]
173          THEN [logTail EXCEPT ![a] = pe.lIndex] ELSE logTail
174      ∧ isLeader′ = IF pe.term > highestBallot[a]
175          THEN [isLeader EXCEPT ![a] = FALSE] ELSE isLeader
176      ∧ (pe.term > highestBallot[a]) ⇒ (isLeader′ = [isLeader EXCEPT ![a] = FALSE])
177      ∧ UNCHANGED ⟨proposedEntries, proposedValues, r1amsgs, r1bmsgs⟩

179  Init ≜ ∧ highestBallot = [a ∈ Acceptor ↦ 0]
180        ∧ isLeader = [a ∈ Acceptor ↦ FALSE]
181        ∧ logTail = [a ∈ Acceptor ↦ −1]
182        ∧ lastIndex = [a ∈ Acceptor ↦ −1]
183        ∧ votes = [a ∈ Acceptor ↦ [i ∈ Index ↦ {}]]
184        ∧ logs = [a ∈ Acceptor ↦ [i ∈ Index ↦ ⟨−1, NoVal⟩]]
185        ∧ raftlogs = [a ∈ Acceptor ↦ [i ∈ Index ↦ ⟨−1, NoVal⟩]]
186        ∧ logBallot = [a ∈ Acceptor ↦ [i ∈ Index ↦ −1]]
187        ∧ proposedEntries = {}
188        ∧ proposedValues = {}
189        ∧ r1amsgs = {}
190        ∧ r1bmsgs = {}
191  ⊢
```

```
192  Here we derive PQL by modifying MultiPaxos
193  VotedFor(a, i, b, v) ≜ ⟨b, v⟩ ∈ votes[a][i]
194  The lease related state functions and subactions
195  LeaseIsActive(p) ≜ ∃ Q ∈ Quorum : ∀ a ∈ Q : leases[a][p] ≥ timer

197  ActiveLeaseHolders ≜ {p ∈ Acceptor : LeaseIsActive(p)}

199  GrantedLeaseHolders(Q) ≜ {p ∈ Acceptor : ∃ a ∈ Q : leases[a][p] ≥ timer}

201    a value can be executable only if all the lease holders have acknowledged this change
202  CanCommitAt(i, b, v) ≜
203      ∃ Q ∈ Quorum :
204          ∧ ∀ a ∈ Q : VotedFor(a, i, b, v)   v is chosen by Q
205          ∧ ∀ p ∈ GrantedLeaseHolders(Q) : VotedFor(p, i, b, v)
206          v is known by every lease holder granted by ∀ a ∈ Q
207          ⇒ v is known by every active lease holder (Proved by the property of Quorum)

209  executable ≜ {ibv ∈ Index × Ballot × Value : CanCommitAt(ibv[1], ibv[2], ibv[3])}

211  ReadAtLocal(a) ≜
212      ∧ LeaseIsActive(a)
213      ∧ logTail[a] = applyIndex[a]   wait for all pending write commands to finish
214      ∧ TRUE   Do local reading (does not change any server state)
215      ∧ UNCHANGED ⟨MP!vars, RQLvars⟩

217  Apply(a, i) ≜
218      ∧ i = applyIndex[a] + 1
219      ∧ CanCommitAt(i, logs[a][i][1], logs[a][i][2])
220      ∧ applyIndex' = [applyIndex EXCEPT ![a] = i]
221      ∧ UNCHANGED MP!vars

223    p grant lease to q
224  GrantLease(p, q) ≜
225      ∧ leases' = [leases EXCEPT ![p][q] = timer + LeaseDuration]
226      ∧ UNCHANGED MP!vars

228  UpdateTimer ≜
229      ∧ timer' = timer + 1
230      ∧ UNCHANGED MP!vars

232  Next ≜  ∨ ∃ a ∈ Acceptor, b ∈ Ballot :
233              IncreaseHighestBallot(a, b) ∧ UNCHANGED RQLvars
234          ∨ ∃ a ∈ Acceptor :
235              Phase1a(a) ∧ UNCHANGED RQLvars
236          ∨ ∃ a ∈ Acceptor, m ∈ r1amsgs :
237              Phase1b(a, m)
238          ∨ ∃ a ∈ Acceptor, S ∈ SUBSET r1bmsgs :
239              BecomeLeader(a, S) ∧ UNCHANGED RQLvars
```

```
240            ∨ ∃ a, x ∈ Acceptor, i1, i ∈ Index, v ∈ Value :
241                ∧ ∨ v.type ≠ "read"
242                  ∨ ¬LeaseIsActive(a)
243                ∧ ProposeEntries(a, i1, i, v)
244                ∧ UNCHANGED RQLvars
245            ∨ ∃ a ∈ Acceptor, pe ∈ proposedEntries :
246                AcceptEntries(a, pe)
247            ∨ ∃ a ∈ Acceptor, i ∈ Index :
248                Apply(a, i)
249            ∨ ∃ a, p ∈ Acceptor :
250                GrantLease(a, p)
251            ∨ ∃ a ∈ Acceptor :
252                ReadAtLocal(a)
253            ∨ UpdateTimer

255   Spec ≜ Init ∧ □[Next]_vars
```

## B.5 Mencius: Coordinated Paxos

The TLA$^+$ Specification starts at the next page.

1 ─────────────── MODULE *CoorPaxos* ───────────────
This is the specification of Coordinated *Paxos*, which is used by Mencius

6 EXTENDS *Integers*
7 $Min(s) \triangleq \text{CHOOSE } x \in s : \forall y \in s : x \leq y$
8 $Max(s) \triangleq \text{CHOOSE } x \in s : \forall y \in s : x \geq y$
9 ────────────────────────────────────────────────

11 CONSTANT *Quorum*, *Value*, *Acceptor*

13 VARIABLE *highestBallot*, *isLeader*, *logTail*,
14     *votes*, *proposedValues*, *logs*, *1amsgs*,
15     *1bmsgs*

17 $Ballot \triangleq Nat$
18 $Index \triangleq Nat$
19 $NoVal \triangleq \text{CHOOSE } v : v \notin Value$
20 $Noop \triangleq \text{CHOOSE } v : v \notin Value \land v \neq NoVal$

22 VARIABLE *skipTags*, *executable*
23 CONSTANT *isDefault*    *isDefault[a] =* True iff a is the default leader
24 ────────────────────────────────────────────────

26 ASSUME $QuorumAssumption \triangleq \land \forall Q \in Quorum : Q \subseteq Quorum$
27     $\land \forall Q1, Q2 \in Quorum : Q1 \cap Q2 \neq \{\}$

29 $VotedFor(a, i, b, v) \triangleq \langle b, v \rangle \in votes[a][i]$

31 $learnedAt(i, b, v) \triangleq \exists Q \in Quorum :$
32     $\forall a \in Q : VotedFor(a, i, b, v)$

34 $learned \triangleq [i \in Index \mapsto \{v \in Value : \exists b \in Ballot : learnedAt(i, b, v)\}]$

35 ────────────────────────────────────────────────
36 $GetHighestBallotEntry(i, lsIn1b) \triangleq$
37     CHOOSE $lsi \in \{\langle ls[1][i], ls[2][i]\rangle : ls \in lsIn1b\} :$
38     $\land \forall logSkiptag \in lsIn1b : lsi[1][1] \geq logSkiptag[1][i][1]$    ballot

40 $UpdateLog(a, lsIn1b, i1, i2) \triangleq$
41     $\land \forall i \in i1 \ldots i2 : logs'[a][i] = GetHighestBallotEntry(i, lsIn1b)[1]$
42     $\land \forall i \in i1 \ldots i2 : skipTags'[a][i] = GetHighestBallotEntry(i, lsIn1b)[2]$
43     $\land \forall x \in Acceptor \setminus \{a\} : \text{UNCHANGED } logs[x]$
44     $\land \forall i \in Index \setminus i1 \ldots i2 : \text{UNCHANGED } logs[a][i]$
45     $\land logs' \in [Acceptor \to [Index \to ((Ballot \cup \{-1\}) \times (Value \cup \{NoVal\}))]]$
46     $\land logTail' =$
47         IF $(i2 > logTail[a])$ THEN $[logTail \text{ EXCEPT } ![a] = i2]$
48             ELSE $logTail$
49 ────────────────────────────────────────────────

50 $IncreaseHighestBallot(a, b) \triangleq$
51     $\land highestBallot[a] < b$

```
52      ∧ highestBallot' = [highestBallot EXCEPT ![a] = b]
53      ∧ isLeader' = [isLeader EXCEPT ![a] = FALSE]
54      ∧ UNCHANGED ⟨logTail, votes, proposedValues, logs, 1amsgs, 1bmsgs⟩

56  Phase1a(a) ≜
57      ∧ ¬isLeader[a]
58      ∧ 1amsgs' = 1amsgs ∪ {[acc ↦ a, bal ↦ highestBallot[a]]}
59      ∧ UNCHANGED ⟨highestBallot, isLeader, logTail, votes, proposedValues, logs, 1bmsgs⟩

61  Phase1b(a, 1amsg) ≜
62      ∧ 1amsg.bal > highestBallot[a]
63      ∧ highestBallot' = [highestBallot EXCEPT ![a] = 1amsg.bal]
64      ∧ isLeader' = [isLeader EXCEPT ![a] = FALSE]
65      ∧ 1bmsgs' = 1bmsgs ∪ {[acc ↦ a, bal ↦ 1amsg.bal,
66          log ↦ logs[a], skipTag ↦ skipTags[a], logTail ↦ logTail[a]]}
67      ∧ UNCHANGED ⟨logTail, votes, proposedValues, logs, 1amsgs⟩

69  BecomeLeader(a, S) ≜
70      ∧ ¬isLeader[a]
71      ∧ ∃ m ∈ S : m.acc = a
72      ∧ ∀ m ∈ S : m.bal = highestBallot[a]
73      ∧ {m.acc : m ∈ S} ∈ Quorum
74      ∧ UpdateLog(a, {⟨m.log, m.skipTag⟩ : m ∈ S}, 0, Max({m.logTail : m ∈ S}))
75      ∧ isLeader' = [isLeader EXCEPT ![a] = TRUE]
76      ∧ UNCHANGED ⟨highestBallot, votes, proposedValues, 1amsgs, 1bmsgs⟩

78  Propose(a, i, v) ≜
79      ∧ isLeader[a]
80      ∧ ∨ logs[a][i][2] = v
81        ∨ logs[a][i][2] = NoVal      ⇒ v is safe at i, b
82      ∧ proposedValues' = proposedValues ∪ {⟨i, highestBallot[a], v, isDefault[a]⟩}
83      ∧ UNCHANGED ⟨highestBallot, isLeader, logTail, votes, logs, 1amsgs, 1bmsgs⟩

85  Accept(a, i, b, v) ≜
86          ∃ default ∈ BOOLEAN :
87          ∧ ⟨i, b, v, default⟩ ∈ proposedValues
88          ∧ b ≥ highestBallot[a]
89          ∧ highestBallot' = [highestBallot EXCEPT ![a] = b]
90          ∧ votes' = [votes EXCEPT ![a][i] = votes[a][i] ∪ {⟨b, v⟩}]
91          ∧ logs' = [logs EXCEPT ![a][i] = ⟨b, v⟩]
92          ∧ (default = TRUE ∧ v = Noop) ⇒
93              (skipTags'[a][i] = TRUE) ∧ executable[a]' = executable[a] ∪ {⟨i, v⟩}
94          ∧ logTail' = IF i > logTail[a] THEN [logTail EXCEPT ![a] = i] ELSE logTail
95          ∧ isLeader' = IF b > highestBallot[a] THEN [isLeader EXCEPT ![a] = FALSE] ELSE isLeader
96          ∧ UNCHANGED ⟨proposedValues, 1amsgs, 1bmsgs⟩
97  ├
```

$$
\begin{array}{rl}
98 & TypeOK \triangleq \wedge highestBallot \in [Acceptor \to Ballot] \\
99 & \wedge isLeader \in [Acceptor \to \text{BOOLEAN}] \\
100 & \wedge logTail \in [Acceptor \to Index \cup \{-1\}] \\
101 & \wedge votes \in [Acceptor \to [Index \to \text{SUBSET}\ (Ballot \times Value)]] \\
102 & \wedge proposedValues \in \text{SUBSET}\ (Index \times Ballot \times Value \times \text{BOOLEAN}) \\
103 & \wedge logs \in [Acceptor \to [Index \to (Ballot \cup \{-1\}) \times (Value \cup \{NoVal\})]] \\
104 & \wedge skipTags \in [Acceptor \to [Index \to \text{BOOLEAN}]] \\
105 & \wedge executable \in [Acceptor \to \text{SUBSET}\ (Index \times (Value \cup \{Noop\}))] \\
106 & \wedge 1amsgs \in \text{SUBSET}\ [acc : Acceptor, bal : Ballot] \\
107 & \wedge 1bmsgs \in \text{SUBSET}\ [acc : Acceptor, bal : Ballot, \\
108 & \quad log : [Index \to (Ballot \cup \{-1\}) \times (Value \cup \{NoVal\})], \\
109 & \quad skipTag : [Acceptor \to [Index \to \text{BOOLEAN}]], logTail : Index \cup \{-1\}]
\end{array}
$$

$$
\begin{array}{rl}
111 & Init \triangleq \wedge highestBallot = [a \in Acceptor \mapsto 0] \\
112 & \wedge isLeader = [a \in Acceptor \mapsto \text{FALSE}] \\
113 & \wedge logTail = [a \in Acceptor \mapsto -1] \\
114 & \wedge votes = [a \in Acceptor \mapsto [i \in Index \mapsto \{\}]] \\
115 & \wedge logs = [a \in Acceptor \mapsto [i \in Index \mapsto \langle -1, NoVal \rangle]] \\
116 & \wedge proposedValues = \{\} \\
117 & \wedge 1amsgs = \{\} \\
118 & \wedge 1bmsgs = \{\} \\
119 & \wedge skipTags = [a \in Acceptor \mapsto [i \in Index \mapsto \text{FALSE}]] \\
120 & \wedge executable = [a \in Acceptor \mapsto \{\}]
\end{array}
$$

$$
\begin{array}{rl}
123 & Next \triangleq \vee \exists\, a \in Acceptor,\ b \in Ballot : \\
124 & \quad IncreaseHighestBallot(a, b) \\
125 & \vee \exists\, a \in Acceptor : \\
126 & \quad Phase1a(a) \\
127 & \vee \exists\, a \in Acceptor,\ m \in 1amsgs : \\
128 & \quad Phase1b(a, m) \\
129 & \vee \exists\, a \in Acceptor,\ S \in \text{SUBSET}\ 1bmsgs : \\
130 & \quad BecomeLeader(a, S) \\
131 & \vee \exists\, a \in Acceptor,\ i \in Index,\ v \in Value : \\
132 & \quad Propose(a, i, v) \\
133 & \vee \exists\, a \in Acceptor,\ pv \in proposedValues : \\
134 & \quad Accept(a, pv[1], pv[2], pv[3])
\end{array}
$$

$$
136\quad vars \triangleq \langle highestBallot, isLeader, logTail, votes, proposedValues, logs, 1amsgs, 1bmsgs \rangle
$$

$$
138\quad Spec \triangleq Init \wedge \Box [Next]_{vars}
$$

## B.6 Raft*-Mencius: Coordinated Raft*

The TLA[+] Specification starts at the next page.

1 ──────────── MODULE *CoorRaft* ────────────

Specification of Coordinated *Raft*, which is ported from *Coorinated* Paxos.

6 EXTENDS *Integers*

8 CONSTANT *Quorum*,
9     *Value*,
10     *Acceptor*

12 VARIABLE *isLeader*,
13     *logTail*,
14     *lastIndex*,
15     *votes*,
16     *raftlogs*,
17     *proposedEntries*,
18     *highestBallot*,
19     *proposedValues*,
20     *logBallot*,
21     *r1amsgs*,
22     *r1bmsgs*

24 $Index \triangleq Nat$
25 $Ballot \triangleq Nat$

27 ASSUME $QuorumAssumption \triangleq\ \land \forall Q \in Quorum : Q \subseteq Quorum \land Q \subseteq Acceptor$
28                $\land \forall Q1, Q2 \in Quorum : Q1 \cap Q2 \neq \{\}$

31 $logs \ \triangleq\ [a \in Acceptor \mapsto [i \in Index \mapsto \langle logBallot[a][i], raftlogs[a][i][2] \rangle]]$

33 $vars \triangleq \langle isLeader, logTail, lastIndex, raftlogs, logs, votes,$
34   $proposedValues, proposedEntries, highestBallot,$
35   $logBallot, r1amsgs, r1bmsgs \rangle$
36 ├──────────────────────────────────────────────
37 $1amsgs \triangleq \{[acc \mapsto m.acc, bal \mapsto m.bal] : m \in r1amsgs\}$
38 $1bmsgs \triangleq r1bmsgs$

40 $MP \triangleq$ INSTANCE *MultiPaxos*

42 $Max(x) \triangleq MP!Max(x)$
43 $NoVal \triangleq MP!NoVal$
44 $Noop \triangleq$ CHOOSE $v : v \notin Value \land v \neq NoVal$

46 VARIABLE *skipTags*, *executable*
47 CONSTANT *isDefault* *isDefault[a]* = True iff a is the default leader
48 ├──────────────────────────────────────────────
49 $TypeOK \triangleq\ \land highestBallot \in [Acceptor \rightarrow Ballot]$
50      $\land isLeader \in [Acceptor \rightarrow \text{BOOLEAN}]$
51      $\land lastIndex \in [Acceptor \rightarrow Index \cup \{-1\}]$

$$
\begin{aligned}
&52 \qquad\qquad\qquad \land \textit{logTail} \in [\textit{Acceptor} \rightarrow \textit{Index} \cup \{-1\}] \\
&53 \qquad\qquad\qquad \land \textit{votes} \in [\textit{Acceptor} \rightarrow [\textit{Index} \rightarrow \text{SUBSET } (\textit{Ballot} \times \textit{Value})]] \\
&54 \qquad\qquad\qquad \land \textit{raftlogs} \in [\textit{Acceptor} \rightarrow \\
&55 \qquad\qquad\qquad\quad [\textit{Index} \rightarrow ((\textit{Ballot} \cup \{-1\}) \times (\textit{Value} \cup \{\textit{NoVal}\}))]] \\
&56 \qquad\qquad\qquad \land \textit{logs} \in [\textit{Acceptor} \rightarrow [\textit{Index} \rightarrow ((\textit{Ballot} \cup \{-1\}) \times (\textit{Value} \cup \{\textit{NoVal}\}))]] \\
&57 \qquad\qquad\qquad \land \textit{proposedEntries} \in \text{SUBSET } [\textit{term} : \textit{Ballot}, \\
&58 \qquad\qquad\qquad\qquad\qquad\qquad\qquad\quad \textit{prevLogTerm} : \textit{Ballot} \cup \{-1\}, \\
&59 \qquad\qquad\qquad\qquad\qquad\qquad\qquad\quad \textit{prevLogIndex} : \textit{Index} \cup \{-1\}, \\
&60 \qquad\qquad\qquad\qquad\qquad\qquad\qquad\quad \textit{lIndex} : \textit{Index} \cup \{-1\}, \\
&61 \qquad\qquad\qquad\qquad\qquad\qquad\qquad\quad \textit{leaderId} : \textit{Acceptor}, \\
&62 \qquad\qquad\qquad\qquad\qquad\qquad\qquad\quad \textit{leaderCommit} : \textit{Index} \cup \{-1\}, \\
&63 \qquad\qquad\qquad\qquad\qquad\qquad\qquad\quad \textit{entries} : [\textit{Index} \rightarrow (\textit{Ballot} \times \textit{Value})] \\
&64 \qquad\qquad\qquad\qquad\qquad\qquad\qquad\quad ] \\
&65 \qquad\qquad\qquad \land \textit{logBallot} \in [\textit{Acceptor} \rightarrow [\textit{Index} \rightarrow \textit{Ballot} \cup \{-1\}]] \\
&66 \qquad\qquad\qquad \land \textit{proposedValues} \in \text{SUBSET } (\textit{Index} \times \textit{Ballot} \times \textit{Value} \times \text{BOOLEAN}) \\
&67 \qquad\qquad\qquad \land \textit{r1amsgs} \in \text{SUBSET } [\textit{acc} : \textit{Acceptor}, \textit{bal} : \textit{Ballot}, \\
&68 \qquad\qquad\qquad\quad \textit{lastTerm} : \textit{Ballot} \cup \{-1\}, \textit{lastIndex} : \textit{Index} \cup \{-1\}] \\
&69 \qquad\qquad\qquad \land \textit{r1bmsgs} \in \text{SUBSET } [\textit{acc} : \textit{Acceptor}, \textit{bal} : \textit{Ballot}, \\
&70 \qquad\qquad\qquad\quad \textit{skipTag} : [\textit{Acceptor} \rightarrow [\textit{Index} \rightarrow \text{BOOLEAN}]], \\
&71 \qquad\qquad\qquad\quad \textit{log} : [\textit{Index} \rightarrow (\textit{Ballot} \cup \{-1\}) \times (\textit{Value} \cup \{\textit{NoVal}\})], \textit{logTail} : \textit{Index} \cup \{-1\}] \\
&72 \qquad\qquad\qquad \land \textit{skipTags} \in [\textit{Acceptor} \rightarrow [\textit{Index} \rightarrow \text{BOOLEAN}]] \\
&73 \qquad\qquad\qquad \land \textit{executable} \in [\textit{Acceptor} \rightarrow \text{SUBSET } (\textit{Index} \times (\textit{Value} \cup \{\textit{Noop}\}))]
\end{aligned}
$$

75 ⊢

$$
\begin{aligned}
&76 \quad \textit{GetHighestBallotEntry}(i, \textit{lsIn1b}) \triangleq \\
&77 \qquad \text{CHOOSE } \textit{lsi} \in \{\langle \textit{ls}[1][i], \textit{ls}[2][i]\rangle : \textit{ls} \quad \in \textit{lsIn1b}\} : \\
&78 \qquad\quad \land \quad \forall \textit{logSkiptag} \in \textit{lsIn1b} : \textit{lsi}[1][1] \geq \textit{logSkiptag}[1][i][1] \quad \boxed{\text{ballot}} \\
\\
&80 \quad \textit{UpdateLog}(a, \textit{lsIn1b}, i1, i2) \triangleq \\
&81 \qquad \land \forall i \in i1 \ldots i2 : \\
&82 \qquad\quad \land \textit{raftlogs}'[a][i] = \langle -1, \textit{GetHighestBallotEntry}(i, \textit{lsIn1b})[1][2]\rangle \\
&83 \qquad\quad \land \textit{logBallot}'[i] = \textit{GetHighestBallotEntry}(i, \textit{lsIn1b})[1][1] \\
&84 \qquad\quad \land \textit{skipTags}'[a] = \textit{GetHighestBallotEntry}(i, \textit{lsIn1b})[2] \\
&85 \qquad \land \forall i \in \textit{Index} \setminus i1 \ldots i2 : \text{UNCHANGED } \langle \textit{raftlogs}[a][i], \textit{logBallot}[a][i]\rangle \\
&86 \qquad \land \forall x \in \textit{Acceptor} \setminus \{a\} : \text{UNCHANGED } \langle \textit{raftlogs}[x], \textit{logBallot}[x]\rangle \\
&87 \qquad \land \textit{raftlogs}' \in [\textit{Acceptor} \rightarrow [\textit{Index} \rightarrow ((\textit{Ballot} \cup \{-1\}) \times (\textit{Value} \cup \{\textit{NoVal}\}))]] \\
&88 \qquad \land \textit{logTail}' = \\
&89 \qquad\quad \text{IF } (i2 > \textit{logTail}[a]) \text{ THEN } [\textit{logTail} \text{ EXCEPT } ![a] = i2] \text{ ELSE } \textit{logTail}
\end{aligned}
$$

90 ⊢

$$
\begin{aligned}
&91 \quad \textit{IncreaseHighestBallot}(a, b) \triangleq \\
&92 \qquad \land \textit{MP}!\textit{IncreaseHighestBallot}(a, b) \\
&93 \qquad \land \text{UNCHANGED } \langle \textit{isLeader}, \textit{lastIndex}, \textit{logTail}, \textit{logs}, \textit{raftlogs}, \\
&94 \qquad\quad \textit{logBallot}, \textit{votes}, \textit{proposedValues}, \textit{proposedEntries}\rangle \\
\\
&96 \quad \textit{Phase1a}(a) \triangleq \\
&97 \qquad \land \quad \textit{r1amsgs}' = \textit{r1amsgs} \cup \{[\textit{acc} \mapsto a,
\end{aligned}
$$

$$
\begin{aligned}
&98 &&bal \mapsto highestBallot[a], \\
&99 &&lastTerm \mapsto \text{IF } lastIndex[a] = -1 \text{ THEN } -1 \\
&100 &&\phantom{lastTerm \mapsto{}} \text{ELSE } raftlogs[a][lastIndex[a]][1], \\
&101 &&lastIndex \mapsto lastIndex[a]]\} \\
&102 &\wedge \text{ UNCHANGED } \langle highestBallot, isLeader, lastIndex, logTail, logs, \\
&103 &&raftlogs, logBallot, votes, proposedValues, proposedEntries, r1bmsgs\rangle
\end{aligned}
$$

105 $Phase1b(a, r1amsg) \triangleq$
106 $\quad\wedge\ r1amsg.bal > highestBallot[a]$
107 $\quad\wedge\ \vee\ lastIndex[a] = -1$
108 $\qquad\vee\ \wedge\ lastIndex[a] \neq -1$
109 $\qquad\quad\ \wedge\ raftlogs[a][lastIndex[a]][1] < r1amsg.lastTerm$
110 $\qquad\vee\ \wedge\ lastIndex[a] \neq -1$
111 $\qquad\quad\ \wedge\ raftlogs[a][lastIndex[a]][1] = r1amsg.lastTerm$
112 $\qquad\quad\ \wedge\ lastIndex[a] \leq r1amsg.lastIndex$
113 $\quad\wedge\ highestBallot' = [highestBallot \text{ EXCEPT } ![a] = r1amsg.bal]$
114 $\quad\wedge\ isLeader' = [isLeader \text{ EXCEPT } ![a] = \text{FALSE}]$
115 $\quad\wedge\ r1bmsgs' = r1bmsgs \cup \{[acc \mapsto a, bal \mapsto r1amsg.bal,$
116 $\qquad log \mapsto logs[a], skipTag \mapsto skipTags[a], logTail \mapsto logTail[a]]\}$
117 $\quad\wedge\ \text{UNCHANGED } \langle lastIndex, logTail, logs, raftlogs, logBallot,$
118 $\qquad votes, proposedValues, proposedEntries, r1amsgs\rangle$

120 $BecomeLeader(a, S) \triangleq$
121 $\quad\wedge\ \exists\, m \in S : m.acc = a$
122 $\quad\wedge\ \forall\, m \in S : m.bal = highestBallot[a]$
123 $\quad\wedge\ \{m.acc : m \in S\} \in Quorum$
124 $\quad\wedge\ \forall\, i \in 0\,..\,lastIndex[a] :$
125 $\qquad \text{UNCHANGED } \langle logBallot[a][i], raftlogs[a][i]\rangle$
126 $\quad\wedge\ UpdateLog(a, \{\langle m.log, m.skipTag\rangle : m \in S\}, lastIndex[a] + 1, Max(\{m.logTail : m \in S\}))$
127 $\quad\wedge\ isLeader' = [isLeader \text{ EXCEPT } ![a] = \text{TRUE}]$
128 $\quad\wedge\ \text{UNCHANGED } \langle highestBallot, lastIndex, votes, proposedValues,$
129 $\qquad proposedEntries, r1bmsgs, r1amsgs\rangle$

131 $ProposeEntries(a, i1, i, v) \triangleq$
132 $\quad\wedge\ isLeader[a]$
133 $\quad\wedge\ i = logTail[a] + 1$
134 $\quad\wedge\ proposedEntries' = proposedEntries\ \cup$
135 $\qquad [term \mapsto highestBallot[a], prevLogTerm \mapsto$
136 $\qquad\quad \text{IF } i1 = 0 \text{ THEN } raftlogs[a][i1 - 1][1] \text{ ELSE } -1,$
137 $\qquad prevLogIndex \mapsto i1 - 1, lIndex \mapsto i, leaderId \mapsto a, isDefault \mapsto isDefault[a],$
138 $\qquad leaderCommit \mapsto -1, entries \mapsto [j \in i1\,..\,i\ \mapsto$
139 $\qquad\quad \text{IF } j = i \text{ THEN } \langle highestBallot[a], v\rangle \text{ ELSE } raftlogs[a][j]]]$
140 $\quad\wedge\ proposedValues' = proposedValues \cup \{ibvs \in (Index \times Ballot \times Value \times \text{BOOLEAN}) :$
141 $\qquad \wedge\ raftlogs[a][ibvs[1]][2] = ibvs[3]$
142 $\qquad \wedge\ ibvs[2] = highestBallot[a]$
143 $\qquad \wedge\ ibvs[1] \in 0\,..\,i$

$$
\begin{aligned}
144 \quad & \wedge ibvs[4] = skipTags[a][ibvs[1]]\} \\
145 \quad & \wedge \text{UNCHANGED } \langle isLeader, lastIndex, logTail, raftlogs, logs, logBallot, \\
146 \quad & \qquad votes, highestBallot, r1amsgs, r1bmsgs \rangle
\end{aligned}
$$

$$
\begin{aligned}
148 \quad & AcceptEntries(a, pe) \triangleq \\
149 \quad & \quad \text{LET } default \triangleq pe.isDefault \text{ IN} \\
150 \quad & \quad \wedge pe.term \geq highestBallot[a] \\
151 \quad & \quad \wedge pe.lIndex \geq lastIndex[a] \\
152 \quad & \quad \wedge (pe.prevLogIndex > -1) \Rightarrow (raftlogs[a][pe.prevLogIndex][1] = pe.prevLogTerm) \\
153 \quad & \quad \wedge \forall i \in 0 \,..\, pe.lIndex : logBallot'[a][i] = pe.term \\
154 \quad & \quad \wedge \forall i \in pe.prevLogIndex + 1 \,..\, pe.lIndex : \\
155 \quad & \qquad \text{LET } v \triangleq pe.entries[i][2] \text{ IN} \\
156 \quad & \qquad \wedge raftlogs'[a][i] = pe.entries[i] \\
157 \quad & \qquad \wedge (default = \text{TRUE} \wedge v = Noop) \Rightarrow \\
158 \quad & \qquad \quad (skipTags'[a][i] = \text{TRUE}) \wedge executable[a]' = executable[a] \cup \{\langle i,\, v\rangle\} \\
159 \quad & \qquad \wedge votes'[a][i] = votes[a][i] \cup \{\langle pe.entries[pe.lIndex][1],\, pe.entries[i][2]\rangle\} \\
160 \quad & \quad \wedge \forall i \in Index \setminus (0 \,..\, pe.lIndex) : \\
161 \quad & \qquad \text{UNCHANGED } \langle raftlogs[a][i], votes[a][i], logBallot[a][i]\rangle \\
162 \quad & \quad \wedge \forall x \in Acceptor \setminus \{a\} : \text{UNCHANGED } \langle raftlogs[x], votes[x], logBallot[x]\rangle \\
163 \quad & \quad \wedge highestBallot' = [highestBallot \text{ EXCEPT } ![a] = pe.term] \\
164 \quad & \quad \wedge lastIndex' = \text{IF } pe.lIndex > lastIndex[a] \\
165 \quad & \qquad \text{THEN } [lastIndex \text{ EXCEPT } ![a] = pe.lIndex] \text{ ELSE } lastIndex \\
166 \quad & \quad \wedge logTail' = \text{IF } pe.lIndex > logTail[a] \\
167 \quad & \qquad \text{THEN } [logTail \text{ EXCEPT } ![a] = pe.lIndex] \text{ ELSE } logTail \\
168 \quad & \quad \wedge isLeader' = \text{IF } pe.term > highestBallot[a] \\
169 \quad & \qquad \text{THEN } [isLeader \text{ EXCEPT } ![a] = \text{FALSE}] \text{ ELSE } isLeader \\
170 \quad & \quad \wedge (pe.term > highestBallot[a]) \Rightarrow (isLeader' = [isLeader \text{ EXCEPT } ![a] = \text{FALSE}]) \\
171 \quad & \quad \wedge \text{UNCHANGED } \langle proposedEntries, proposedValues, r1amsgs, r1bmsgs\rangle
\end{aligned}
$$

$$
\begin{aligned}
173 \quad & Init \triangleq \wedge highestBallot = [a \in Acceptor \mapsto 0] \\
174 \quad & \qquad \wedge isLeader = [a \in Acceptor \mapsto \text{FALSE}] \\
175 \quad & \qquad \wedge logTail = [a \in Acceptor \mapsto -1] \\
176 \quad & \qquad \wedge lastIndex = [a \in Acceptor \mapsto -1] \\
177 \quad & \qquad \wedge votes = [a \in Acceptor \mapsto [i \in Index \mapsto \{\}]] \\
178 \quad & \qquad \wedge logs = [a \in Acceptor \mapsto [i \in Index \mapsto \langle -1, NoVal\rangle]] \\
179 \quad & \qquad \wedge raftlogs = [a \in Acceptor \mapsto [i \in Index \mapsto \langle -1, NoVal\rangle]] \\
180 \quad & \qquad \wedge logBallot = [a \in Acceptor \mapsto [i \in Index \mapsto -1]] \\
181 \quad & \qquad \wedge proposedEntries = \{\} \\
182 \quad & \qquad \wedge proposedValues = \{\} \\
183 \quad & \qquad \wedge r1amsgs = \{\} \\
184 \quad & \qquad \wedge r1bmsgs = \{\} \\
185 \quad & \qquad \wedge skipTags = [a \in Acceptor \mapsto [i \in Index \mapsto \text{FALSE}]] \\
186 \quad & \qquad \wedge executable = [a \in Acceptor \mapsto \{\}]
\end{aligned}
$$

$$
\begin{aligned}
188 \quad & Next \triangleq \vee \exists\, a \in Acceptor,\, b \in Ballot : \\
189 \quad & \qquad IncreaseHighestBallot(a, b)
\end{aligned}
$$

```
190              ∨ ∃ a ∈ Acceptor :
191                    Phase1a(a)
192              ∨ ∃ a ∈ Acceptor, m ∈ r1amsgs :
193                    Phase1b(a, m)
194              ∨ ∃ a ∈ Acceptor, S ∈ SUBSET r1bmsgs :
195                    BecomeLeader(a, S)
196              ∨ ∃ a, x ∈ Acceptor, i1, i ∈ Index, v ∈ Value :
197                    ProposeEntries(a, i1, i, v)
198              ∨ ∃ a ∈ Acceptor, pe ∈ proposedEntries :
199                    AcceptEntries(a, pe)

201   Spec  ≜  Init ∧ □[Next]_vars
202
```

## C   PROOF OF THE REFINEMENT MAPPING FROM RAFT* TO PAXOS

The mapping between the variables of Raft* and Paxos is:

| Paxos | Raft |
|---|---|
| ballot | currentTerm |
| logEntry.index | logEntry.index |
| logEntry.accBallot | lastLogEntry.Term |
| logEntry.value | logEntry.value |
| phase1Succeeded | isLeader |
| chosenSet | log[0,commitIndex] |
| msg1a | msgRequestVote |
| msg1b | msgRequestVoteOK |
| msg2a | msgAppend |
| msg2b | msgAppendOK |

Note: logEntry stands for each instance of Paxos and Raft* Note: The use of "unchosen" in Phase 1 of Paxos is an optimization that omit transfering data that has been chosen. We do not use this optimization in Raft* so we omit this optimization in the proof. So MultiPaxos will transfer entire logs duing Phase 1 in this proof.

To prove the refinement relationship between Raft* and Paxos, it suffices to prove that when Raft* runs, the variables constructed by the mapping (the left column of the above table) always satisfies the rules of Paxos.

Proof. sffamily

THEORY Raft* is a refinement of Paxos

1. The initialization of Raft* implies Paxos's initial states

    Proof: By definition of Raft* and Paxos. We hereby declare the initial state of Raft* and Paxos

    1.1 The initial state of Raft* is:

        for each node: currentTerm = 0; isLeader = False; lastIndex = -1; log = $\bot$;commitIndex = -1;

    1.2 The initial state of Paxos should satisfy:

        for each node: ballot = 0; phase1Succeeded = False; log = $\bot$; chosenSet = $\emptyset$;

    1.3 At the beginning, there is no message existing in the system.

    1.4 Q.E.D.

        Proof: Paxos.ballot is Raft*.currentTerm; Paxos.phase1Succeeded is Raft*.isLeader; Both logs are emtpy.

2. In the specification of Raft*, each function(step) can imply a function of Paxos.

    2.1 RequestVote(s) => Phase1a(s)

        Proof: By definition of RequestVote and Phase1a, both functions increases ballot(term) by 1. And wecan construct a message "prepare" with each message "requestVote". So the state transfer function RequestVote(s) satisfies the specification of Phase1a(s).

    2.2 ReceiveVote(s) => Phase1b(s)

        3.1 It suffices to assume the function ReceiveVote(s) is executed, and prove that every change on each variable satisfies the specifications of Phase1b(s).

      Proof: By the definition of state transfer

  So we will assume that ReceiveVote(s) is executed. Now it suffices to prove the next several points:

  3.2 s receives <"prepare", b> and b > s.ballot

      Proof: By definition of ReceiveVote

  3.3 s.ballot is set to b

      Proof: By definition of ReceiveVote

  3.4 s.Phase1Succeeded is set to false

      Proof: By definition of ReceiveVote

  3.5 s sends a message <"prepareOK", s.ballot, s.log>

      Proof: By definition of ReceiveVote, and we can map each "requestVoteOK" message to one "prepareOK" message, the ballot in the message is the currentTerm in "requestVoteOK" message. the log entries in "prepareOK" message should be the current log of s. Note that Raft* does not need to send the entire log because of the log matching property, but without loss of generality, we can still assume Raft* includes the full log in the message.

  3.6 all other variables do not change during this state transfer

      Proof: By definition of the two functions

  3.7 Q.E.D.

      Proof: By 3.1-3.6

2.3 BecomeLeader(s) => Phase1Succeed(s)

  3.1 It suffices to assume BeconeLeader(s) is executed, and prove that the state transfer satisfies function Phase1Succeed.

      Proof: By the definition of state transfer

  So we will assume that BecomeLeader(s) is executed. Then we will have:

  3.2 s reveives <"prepareOK", b, instances> from f+1 acceptors with the same b, and b == s.ballot.

      Proof: By definition of the two functions

  3.3 let "end" be the largest id of received instances:

    for i in 0..end: s.instances[i] = safeEntry(received instances with id i)

    4.1 we can cut the log of s into two parts: the first part is 0..s.lastIndex, and the second part is s.lastIndex+1..end

    4.2 The first part: s.instances[0,s.lastIndex] is already the safeEntries of Paxos algorithm

    4.3 The second part: s.instances[s.lastIndex+1..end]

  3.4 s.phase1Succeeded = true

      Proof: By definition of the two functions

  3.5 all other variables do not change during this state transfer

      Proof: By definition of the two functions

  3.6 Q.E.D.

      Proof: By 3.1-3.5

2.4 AppendEntries(s, vals, prev) => Phase2a(s, i, v)

  3.1 It suffices to assume AppendEntries(s, vals, prev) is executed, and prove that the state transfer satisfies a continuous execution of a sequence of function Phase2a. (stuttering, i.e. one AppendEntries step is mapped to several Phase2a steps, because "AppendEntries" batches a sequence of values but Phase2a only send one value at a time.)

  So we will assume that AppendEntries(s, vals, prev) is executed. Then for each i in prev+1..prev+|vals|, we will have:

  3.2 s.phase1Succeeded == true

  3.3 s.instances[i].val == v || s.instances[i].val == Empty

  3.4 s sends message <"accept", i, v, s.ballot>

  3.5 all other variables do not change during this state transfer

  3.6 Q.E.D.

   Proof: By 3.1-3.5

 2.5 ReceiveAppend(s) => Phase2b(s)

  3.1 It suffices to assume ReceiveAppend(s) is executed, and prove that the state transfer satisfies a continuous execution of a sequence of function Phase2b. (stuttering, i.e. one ReceiveAppend step is mapped to several Phase2b steps, because "ReceiveAppend" batches a sequence of values but Phase2b only handle one value at a time.)

  So we will assume that one ReceiveAppend(s) is executed. Then for each i in prev+1..prev+|ents|, we will have:

  3.2 s receives <"accept", i, v, b> and b >= s.ballot

  3.3 if b > s.ballot then s.phase1Succeeded = false; s.ballot = b

  3.4 s.instances[i].bal = b and s.instances[i].val = v

  3.5 s sends message <"acceptOK", i, v, b>

  3.6 all other variables do not change during this state transfer

  3.7 Q.E.D.

   Proof: By 3.1-3.6

 2.6 LeaderLearn(s) => Learn(s)

  3.1 It suffices to assume the function LeaderLearn(s) is executed, and prove that every change on each variable satisfies the specifications of Learn(s).

   Proof: By the definition of state transfer

  So we will assume that LeaderLearn(s) is executed. Now it suffices to prove the next several points:

  3.2 s receives same <"acceptOK", i, v, b> from f+1 acceptors

  3.3 s.instances[i].bal = b

  3.4 s.instances[i].val = v

  3.5 s.instances[i] is added to s.chosenSet

  3.6 all other variables do not change during this state transfer

  3.7 Q.E.D.

   Proof: By 3.1-3.6

 2.7 Q.E.D.

   Proof: By 2.1 - 2,6, after each function of Raft*, the changes of the constructed variables satisfy one type of corresponding state transfer function of Paxos.

3. Q.E.D.

  Proof: By 1 and 2, view Paxos and Raft* as specifications of state machines, the initial states are the same, and each step of Raft* implies a step of Paxos. So we can say the behavior of Raft* satisfies the specification of Paxos, i.e. Raft* is a refinement of Paxos.

 □